\documentclass[fleqn,usenatbib]{mnras}

\usepackage{newtxtext,newtxmath}

\usepackage[T1]{fontenc}

\DeclareRobustCommand{\VAN}[3]{#2}
\let\VANthebibliography\thebibliography
\def\thebibliography{\DeclareRobustCommand{\VAN}[3]{##3}\VANthebibliography}

\usepackage{graphicx}	
\usepackage{amsmath}	

\newcommand{\kms}{\,km\,s$^{-1}$}

\title[Short-period PCEBs identified from SDSS and LAMOST]{Short-period Post-Common Envelope Binaries with Balmer Emission from SDSS and LAMOST Based on ZTF Photometric Data}

\author[Li \& Zhang]{
 Li Lifang,$^{1,2,3}$\thanks{E-mail:llf@ynao.ac.cn or Zhangfh@ynao.ac.cn}
and Zhang Fenghui,$^{1,2,3}$
\\
$^{1}$Yunnan Observatories, Chinese Academy of Sciences, 650216, Kunming, P.R. China\\
$^{2}$Center for Astronomical Mega-Science, Chinese Academy of Sciences, 20A Datun Road, Beijing, 100012, P.R. China\\
$^{3}$Key Laboratory for the Structure and Evolution of Celestial Objects, Chinese Academy of Sciences, 650216, Kunming, P.R. China
}

\date{Accepted yymmdd. Received yymmdd; in original form yymmdd}

\pubyear{2023}

\begin{document}
\label{firstpage}
\pagerange{\pageref{firstpage}--\pageref{lastpage}}
\maketitle

\begin{abstract}
We present here 55 short period PCEBs containing a hot WD and
a low-mass MS. Based on the photometric data from ZTF DR19, the light curves  are analyzed
for about 200 WDMS binaries with emission line(s) identified from SDSS or LAMOST spectra, in which 55 WDMS binaries are found to
exhibit variability in their luminosities with a short period and are thus short-period binaries (i.e. PCEBs).
In addition, it is found that the orbital periods of these PCEBs
locate in a range from 2.2643 to 81.1526 hours. However, only 6
short-period PCEBs are newly discovered and the orbital periods of 19 PCEBs are improved in this work.
Meanwhile, it is found that three objects are newly discovered eclipsing PCEBs, and a object
(i.e. SDSS J1541) might be the short-period PCEB with a late M-type star or a brown dwarf companion
based on the analysis of its spectral energy distribution.
At last, the mechanism(s) being responsible for the emission features
in the spectra of these PCEBs are discussed, the emission features arising in their optical spectra might be
caused by the stellar activity or an irradiated component owing to a hot white dwarf companion because most of them contain a white dwarf
with an effective temperature higher than $\sim$10,000 K.
\end{abstract}




\section{Introduction}
The typical final products of stellar evolution for about 98 per cent of main sequence (MS) stars are white dwarfs (WDs), because
their masses are too lower to ignite He, C / O, or ONe so that nuclear reactions ceases, leaving a degenerate core
of He, C / O, or ONe after their envelopes are lost owing to stellar wind or mass transfer interactions \citep{Brown2011,Gil2001}. 
However, in some cases, He is ignited if no mass transfer interactions take place or they take place in AGB. Meeanwhile, He in the core is also probably ignited if the rapid mass transfer interactions take place near the tip of RGB then leave a hot subdwarf ($\sim 0.5 M_{\sun}$) with a  shell mass of $\la0.02M_{\sun}$ \citep{Heber1986,Han2002,Lei2013}.
Close compact binaries are usually thought to be the products of
common envelope (CE) evolution \citep{Paczynski1976} which is a result of the unstable mass transfer
depended on the mass ratio of binary systems \citep{Nebot2011,Rappaport2017} once the massive star is
on the giant branch (GB) or asymptotic giant branch (AGB) with a radius of 
about 100 $R_{\sun}$ \citep{Webbink1984,Willems2004}  and about 25 per cent binaries would undergo CE phase \citep{Willems2004,Parsons2017}. During
the CE phase, the orbital energy of the donor's core and its companion
is rapidly injected into the CE owning to differential rotation, decreasing
the separation between two components of binary system. If the orbital energy is
enough to eject the envelope before merger, exposing a post-common envelope binary (PCEB), consisting of a WD
and a companion, generally an MS star \citep{Webbink1984,Kao2016,Rappaport2017}.

The CE evolution phase plays an important role in many evolutionary pathways leading
to the formation of compact objects in short period binaries,  such as millisecond pulsars, X-ray binaries, CVs, double WDs, double neutron stars, and strongly
magnetized WDs , or even double black holes \citep{Taam2000,Nebot2009,Rebassa2007,Rebassa2012,Nordhaus2011,Abbott2016}. Although the main features in CE phase had been sketched by \citet{Paczynski1976} more than 40
years ago, then a lot of studies on the formation of PCEBs had been carried out by many
investigators \citep[e.g.][]{Webbink1984,Iben1993,Willems2004,Davis2008,Politano2010}, our understanding on the CE
evolution is still very poor \citep{Politano2007,Nebot2009,Rebassa2012}.  Any significant progress in our understanding of the CE evolution certainly requires both unrelenting
theoretical efforts and innovative observational input \citep{Schreiber2008}. Since the CE phase usually lasts a very short timescale  \citep[$\la$1000 yrs,][]{Hjellming1991,Taam2000,Webbink2008} and
is thus virtually impossible to observe directly. The heavy responsibility of restricting the CE phase falls on objects that have most probably undergone a CE phase
in their past \citep{van2007}.  Among all PCEBs, those containing a WD and a MS star, i.e. WDMS binaries, represent the most
promising population for deriving such observational constraints since they are
very common population of PCEBs \citep{Nebot2009,Nebot2011,Rebassa2012}.

Up to now, about 5,000 WDMS binaries have been discovered by various investigators \citep{Silvestri2006,Heller2009,
Liu2012,Rebassa2013,Ren2013,Li2014,Guo2015} based on SDSS \citep{York2000} or LAMOST spectra
since \citet{Raymond2003} and \citet{Schreiber2003} first attempted to study the WDMS binaries in SDSS.  Meanwhile, near 600 WDMS binaries had been identified
based on the location within the H-R diagram based on Gaia photometric magnitudes \citep{Rebassa2021,Rebassa2023}.
However, the orbital periods of only a small portion of them had been determined by various investigators
through radial velocity observations \citep[e.g.][]{Rebassa2008,Pyrzas2009,Nebot2011,Parsons2021} or photometric data of CRTS, PTF 
or ZTF \citep[e.g.][]{Parsons2013,Drake2014,Parsons2015,Parsons2017,Chen2020,Brown2022} although the number of WDMS binaries
with determined orbital periods is gradually increasing. In the
discovered PCEBs, some of them were found to contain a low mass He WD and a MS companion with $M\la 0.2 M_{\sun}$
\citep{Nebot2011,Luhman2011,Xu2015,Farihi2017,Rappaport2017}, however the oldest globular clusters in the Galactic halo are
currently producing $\sim$ 0.53$M_{\sun}$ WDs from MS progenitors with $M\la 0.8 M_{\sun}$ \citep{Kalirai2009,Brown2011,Rebassa2011}.
Therefore, the low-mass He WDs in these binaries should be formed from the enhanced mass loss from post-MS stars without
reaching asymptotic branch (AGB) and without ever igniting He in interacting binaries \citep{Webbink1984,Nebot2009,Brown2011}.
Although the formation scenario for such binaries had been proposed by \citet{Nelemans1998}, however, we
do not know how is the relatively dense envelope of their progenitors with mass more than 1.0 $M_{\sun}$
expelled by a low-mass companion with a relatively low orbital energy.  In fact, there are a fair number of single low-mass WDs
that show neither variability in their radial velocities nor infrared excess \citep{Rebassa2013}.  The existence
of single low-mass WDs had been explained by many investigators \citep[e.g.][]{Han2002,Killic2007,Justham2009}. Therefore, it is necessary to find
more PCEBs for limiting the results of CE evolution.

In this work, the light curves from ZTF DR19 are analyzed for about 200 WDMS binaries with emission lines identified from SDSS or LAMOST, 
then 55 WDMS binaries are found to be PCEBs with a short period located in a range  from 2.2643 to 81.1526 hours. Among these PCEBs , only 6 are
newly discovered, and the orbital periods of 19 PCEBs are improved. In addition, three new eclipsing PCEBs are found out from these PCEBs based on
their light curves.
The analysis of the photometric data collected from
ZTF survey is presented in Sect. 2. In Sect. 3, we discuss the results and draw our conclusions,

\section{Photometric data analysis}

Although many WDMS binaries ({about 6,000) were identified based on their spectra from SDSS and LAMOST or by their location in
H-R diagram based on the $Gaia$ photometric observations, however
only a small portion of them were found to be the short-period PCEBs through the analysis of their radial velocities or light curves. We
have checked the spectra of some known short-period PCEBs \citep{Rebassa2008,Parsons2013} based on SDSS and/or LAMOST observations carefully, it is found that 
most of them exhibit emission features in their optical spectra. This implies that short-period PCEBs might be easily discovered from WDMS binaries with the emission line(s),
therefore we attempt to find more PCEBs from about 200 WDMS binaries with emission line(s) at Balmer series \citep{Silvestri2007,Rebassa2010,Liu2012,Li2014,Guo2015,Kepler2015}
through detecting the variability in their luminosities based on the analysis of the photometric data
from ZTF survey \citep{Graham2019,Masci2019}. As a result, it is found that 55 binary systems exhibit variability in their luminosities through the analysis of their light curves,
and their orbital periods are found to locate in a range from 2.2643 to 81.1526 hours (listed in Table~\ref{tab:lc_table}), implying that they are short-period PCEBs. Then we match them with Simbad database, it is found that 6 short-period PCEBs are newly discovered and the orbital periods have been improved for 19 PCEBs.  A detailed comparison between the orbital periods obtained by us and their known ones for 49 known PCEBs is shown in Fig.~\ref{fig:spec_period}. As seen from Fig.~\ref{fig:spec_period}, the orbital periods derived in this work are consistent with the known periods for 30 short-period PCEBs except that the orbital periods derived by us for 8 PCEBs (displayed by squares ) are different from the upper limits of their orbital periods estimated by \citet{Morgan2012} based on several radial velocity (RV) measurements. In addition, the orbital periods obtained for 5 PCEBs (indicated by triangles) by us are only a half of their known ones resulted by an assumption that each full light curve should exhibit two maxima and two minima because of ellipsoidal effect. However we do not find any difference between two maxima or two minima in their light curves when their known periods are used to calculate the phases for their light curves, implying that our results for these PCEBs might be correct, however they must be confirmed by the radial velocity observations in the future. The newly discovered PCEBs, together with 14 PCEBs with an improved orbital period, are as the followings:

\subsection{Short-period PCEBs with Balmer emission line(s)}

\subsubsection{SDSS J0029}
SDSS J002926.82+252553.90 (hereafter SDSS J0029) was first identified as a DA WD by \citet{Skiff2009}, then its atmospheric and physical parameters
were derived to be $T_{\rm eff}=19,148(173)$ K, ${\rm log}g=7.58(3)$ and $M=0.459(7)M_{\sun}$ by \citet{Kepler2019} based on the spectrum from SDSS DR14.
This suggests that SDSS J0029 is a hot low-mass He-WD which  might need a friend to be a short-period PCEB \citep{Marsh1995}.

Its spectrum from SDSS indicates that this WD should be companied by a cool M-type star with the Na I$\lambda \lambda$ 8183.27 and 8194.81 absorption
doublet (see Fig.~\ref{fig:spec_figure}). Meanwhile, it is found in Fig.~\ref{fig:spec_figure} that the H$\alpha$ emission presents in its spectrum.  The emission feature may arise due to wind accretion, magnetic activity and/or irradiation because of a hot WD companion, then the reflection effect or star spot (due to magnetic activity) will provide a favorable opportunity for detecting the variability in its luminosity. Thereupon, we collect 724 data points in $g$-band,
975 data points in $r$-band and 133 data points in $i$-band from ZTF DR19 for this object, it is found that the light curves in $g$ and $r$-bands exhibit a large scatter which
is much larger than their observational uncertainties. This implies that the luminosity of this object might be changed with the time, therefore after some scattered data points are
removed, the light curves (720 data points in $g$-band, 953 data points in $r$-band and 131 data points in $i$-band) of this
object are analyzed through a code named {\it Period04} \citep{Lenz2005}. Its orbital period is derived to be  0.12165200 days in $g$-band, 0.12165254 days in $r$-band or 0.12165232 days in $i$-band and the errors ($3\sigma$) of three periods are derived to be $5.3\, 10^{-7}$, $2.7\, 10^{-7}$ and $3.3\, 10^{-7}$ days, respectively. The periodograms
indicating the orbital periods for SDSS J0029 are shown in the left panels of Fig.~\ref{fig:ZTF0029-power_figure}. In general, the amplitude of the luminosity variation because of reflection effect or spot activity in $r$-band is relatively larger than that in $g$-band for the PCEBs with a hot WD and an M-type star and thus the orbital period determined by photometric data in $r$-band is more accurate than that derived from the observations in $g$-band. Therefore, the results on the orbital period derived from $r$-band data are listed in Table~\ref{tab:lc_table}.  At last, the phase for each data point is calculated based on an orbital period of 0.12165254 days, the phase-folded light curves are shown in the right top panel of Fig.~\ref{fig:ZTF0029-power_figure}. Meanwhile, as seen from the left panels of Fig.~\ref{fig:ZTF0029-power_figure}, there might be other possible orbital periods for this object that might not be ruled out, since the power is similar for all peaks. Therefore, two possible orbital periods indicated by two peaks near the main peak (the highest one) are derived to be 0.1084267(1) and 0.1385524(2) days respectively based on the $r$-band data, and the corresponding light curves are plotted in the medium and bottom panels of the right panels of Fig.~\ref{fig:ZTF0029-power_figure}. As seen from the right panels of Fig.~\ref{fig:ZTF0029-power_figure}, It is difficult to directly find the difference in the light curves based on the different periods. In order to obtain the orbital period for this object, we combine 953 data points into 50 normal points in $r$-band observations for three cases mentioned above and calculate their $\chi^2$-values (where $\chi^2=\frac{1}{N}\sum\limits^N_i\frac{(r_{ i}-r_{{\rm nor},i})^2}{\sigma_{i}^2}$, in which $r_i$ is the magnitude of the $i$th data point in $r$-band observations, $r_{{\rm nor},i}$ is a magnitude obtained through interpolation in the magnitudes of the normal points based on the phase of the $i$th data point and $\sigma_i$ the observation uncertainty of the $i$th data point) which are listed in Table~\ref{tab:chi2_table}. It is found that the $\chi^2$-value based on the period indicated by the main peak is a smallest one. This implies that the light curves based on the orbital period indicated by the main peak are most smoothed ones and display fewer scatter data points based on the same photometric data. Therefore, the orbital period of this object should be the one indicated by the main peak in the periodograms of SDSS J0029. This suggests that SDSS J0029 is a short-period PCEBs with an orbital period of about 2.9197 hours.

\begin{figure}
	\includegraphics[width=\columnwidth]{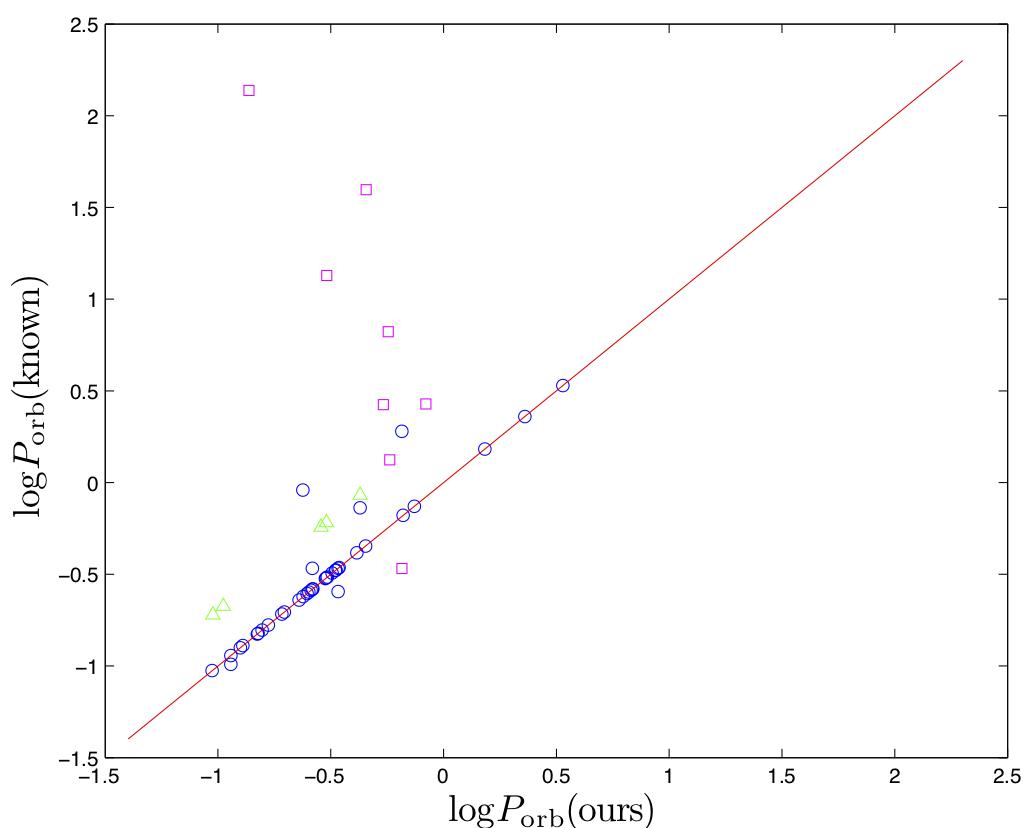}
    \caption{A comparison between the known orbital periods and the ones obtained by us for 49 known short-period PCEBs with hydrogen emission line(s) from SDSS or LAMOST.
     Solid line represents that the orbital periods obtained by us are consistent with the known ones. Squares represent 8 PCEBs with an upper limit in their orbital periods \citep{Morgan2012}, triangles indicate 5 PCEBs with known orbital periods that are twice those given by us and open dots represent other short-period PCEBs.}
    \label{fig:spec_period}
\end{figure}

\begin{figure*}
	\includegraphics[width=18.cm,height=18.cm]{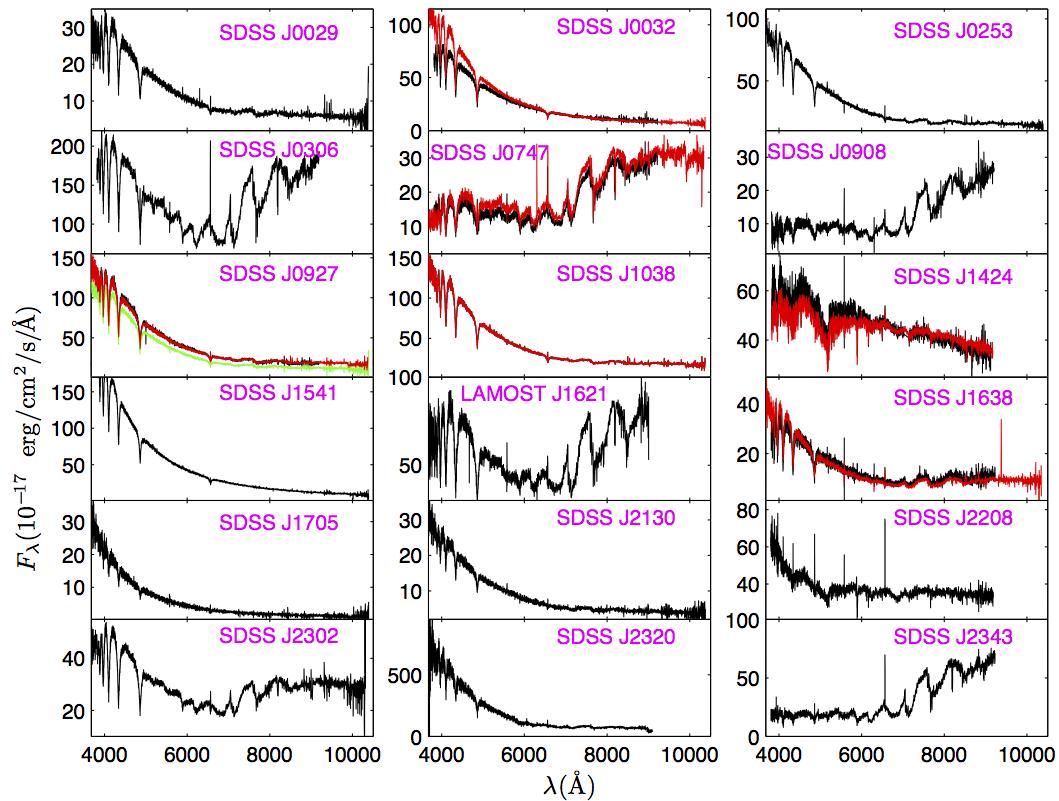}
    \caption{The optical spectra for short-period PCEBs with hydrogen emission line(s) from SDSS or LAMOST. The different colours indicate the different available spectra.}
    \label{fig:spec_figure}
\end{figure*}

\subsubsection{SDSS J0032}

SDSS J003221.87+073934.50 (SDSS J0032)  was first classified as a PCEB by \citet{Schreiber2010} who
gave the atmospheric and physical parameters for this object as the followings: $T_{\rm eff}=21,045$ K, $M_{\rm WD}=0.38M_{\sun}$,
$M_{\rm s}=0.431M_{\sun}$, $d=398$ pc, and a peak to peak radial velocity variation of 298.70 \kms. Then these parameters
were investigated again by other investigators \citep[e.g.][]{Rebassa2010,Girven2011,Kepler2019}.

The WD component of this binary system exhibits a large variability in its radial velocities, implying that SDSS J0032 should be a short-period
PCEB. Therefore, we collect the photometric data from
ZTF DR19 and find that this object indeed shows an evident change in its luminosity. Using the same method as that used for SDSS J0029, its orbital period is
derived to be 0.1539950(14) days and 0.1539945(27) days based on the $r$ and $i$-band data, respectively and the results derived from the $r$-band data are also listed in Table~\ref{tab:lc_table}. Since the periodogram based on $i$-band data displays a similar peak distribution to that based on $r$-band data, so only one periodogram indicating the orbital period for this object based on the $r$-band data is shown in Fig.~\ref{fig:ZTF00321_figure}. We also calculate the $\chi^2$-values for the main peak and a peak with a frequency of 5.4910388 ${\rm d}^{-1}$ ($P_{\rm orb}=0.1821151$ days) and a similar power to the main one  which are also listed in Table~\ref{tab:chi2_table}. It is found in Table~\ref{tab:chi2_table} that the $\chi^2$-value based on a period implied by the main peak is the smaller one. This suggests that the period indicated by the main peak should be its orbital period and its light curves based on an orbital period of 0.15399450
days are shown in Fig.~\ref{fig:ZTF0032_figure}(a). It is found in Fig.~\ref{fig:ZTF0032_figure}(a) that this object is indeed a short-period PCEB, and the periodic
variation in its luminosity might be caused by the reflection effect or spot activity of its MS component.

\begin{figure*}
	\includegraphics[width=8.5cm, height=8.5cm]{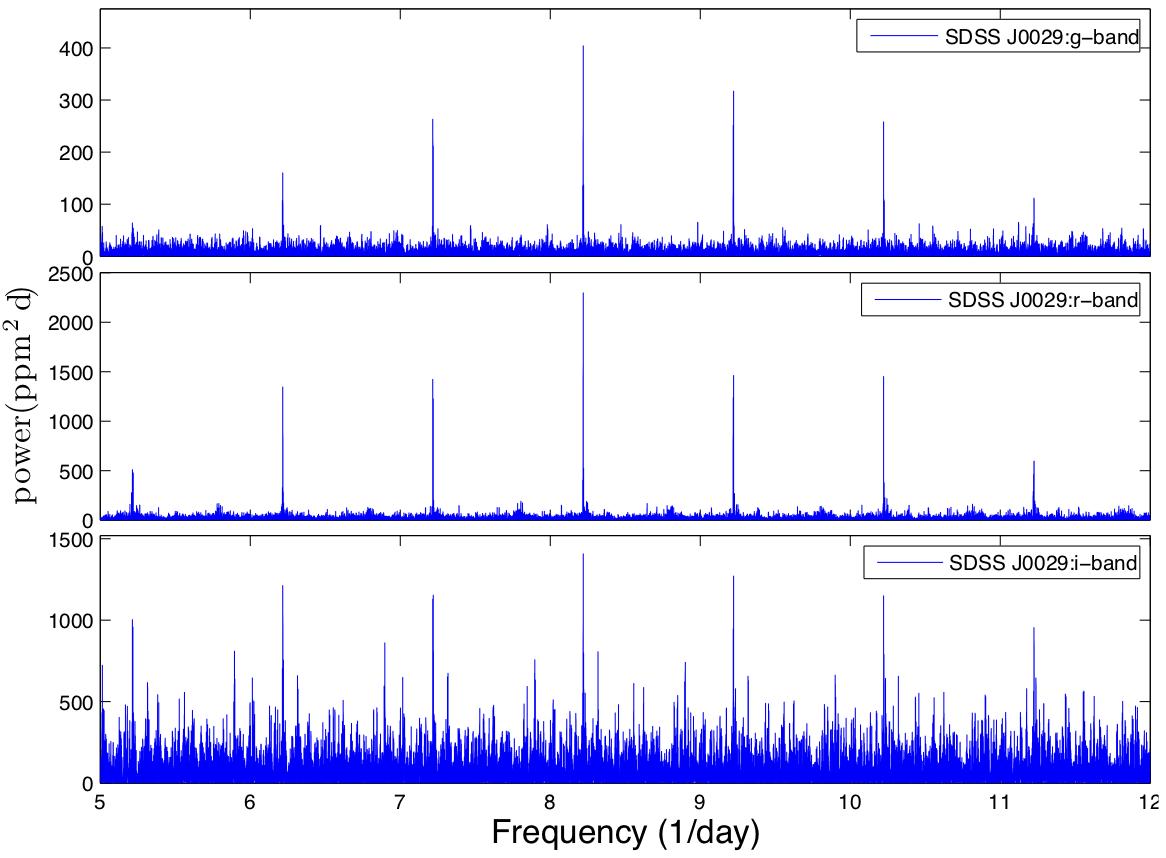}
	\includegraphics[width=8.5cm, height=8.5cm]{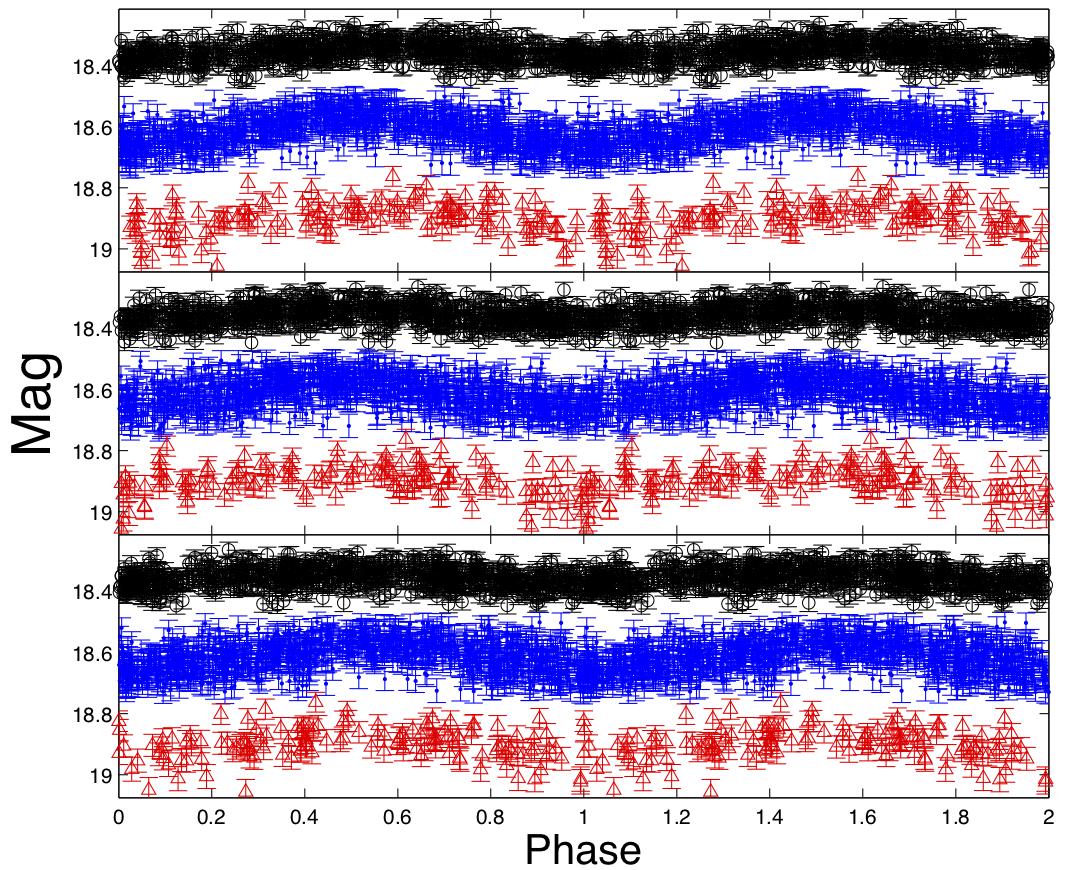}
	    \caption{The periodograms indicating orbital periods and the phase folded light curves for SDSS J0029 based on the ZTF $g$, $r$ and $i$-band photometric observations, respectively.}
    \label{fig:ZTF0029-power_figure}
\end{figure*}

\begin{table*}
	\centering
	\caption{The results on analysis of the light curves for 55 PCEBs ({\bf the first 25 objects are newly discovered or period corrected PCEBs}).}
	\label{tab:lc_table}
	\begin{tabular}{lccccccccccr} 
		\hline
		Stars & RA&Dec&$T_{\rm WD}$&${\rm log}g_{WD}$&$M_{\rm WD}$&Type&$P_{\rm orb}$ & Errs($3\sigma$) & Amplitude&$P_{\rm orb}^{\rm known}$&  Notes \\
		 &(J2000) &(J2000)&(K)&&($M_{\sun}$)&&(days)  & ($10^{-5}$ days)    &  (mag)&(days) &\\ \hline
		SDSS J0029 &7.36183&+25.43177&19148&7.578&0.458&  DA/Me&0.12165254 &0.027 &0.0479&$-$&(1)\\ 
		SDSS J0032 & 8.09105&+7.65965&21045&7.43&0.38&DA/Me&  0.1539950&0.14 & 0.0296&$-$&(2)\\ 
		SDSS J0253 & 43.25668&$-$1.50177&24726&7.61&0.45&DA/Me&0.4264303& 0.15 & 0.0374&0.72904$^{\rm a}$&(4)\\ 
		SDSS J0306 & 46.52996&$-$0.52066&21640&7.22&0.39&DA/Me&0.541158& 5.4 & 0.0217&2.66$^{\rm b}$&(5)\\ 
		SDSS J0747 &116.87735&+43.06765& 15964&8.37&0.85&DA/Me&0.5781084 & 1.7& 0.0395&1.33$^{\rm b}$&(2)\\ 
		SDSS J0908 &137.19754&+61.52831 &18160&8.25&0.83&DA/Me&0. 1372305&  0.07 &0.1974&137.43$^{\rm b}$&(6)\\  
		SDSS J0927 &141.80006&+28.77479& 22037&7.80&0.52&DA/Me& 0.3036211 &0.47 &0.0190&13.46$^{\rm b}$ &(2)\\ 
		SDSS J1038&159.65513&+1.84959&34332&7.56&0.47&DA/Me&0.835045&3.5&0.0125&2.68$^{\rm b}$&(7)\\ 
		SDSS J1424& 216.73090&44.54029& $-$ &$-$&$-$&WD/K7e &0.3549873 &0.29&0. 0141 &              $-$&(2)\\
		SDSS J1541&235.33267&+12.15400&25369&7.44&0.44&DA/Me&0.1140253 & 0.10&0.0599&0.10232$^{\rm c}$&(8)\\ 
		LAMOST J1621&245.30287&+41.30261&14540&7.88&0.55&DA/Me&3.198935&9.1&0.0080&$-$&(3)\\ 
		SDSS J1638&249.60319&+29.45031&19000&7.5&0.39&DA/Me&0.454168&1.6&0.0230&39.49$^{\rm b}$&(9)\\ 
		SDSS J1705&256.32455&+33.75211&34592&7.24&0.44&DA/Me&0.3405451&0.81&0.0987&0.254066$^{\rm c}$&(2)\\
		SDSS J2130&322.58247&+6.20127&34131&7.73&0.53&DA/Me&0.2621132&0.12&0.1112&0.341167$^{\rm e}$& (1)\\ 
		SDSS 2208&332.20414&+12.36241&86726&9.23&1.32&DA/Me&0.654242&1.4&0.0173&0.34$^{\rm b}$,1.903$^{\rm d}$&(2)\\ 
		SDSS J2302&345.51029&$-$00.15844&19416&8.02&0.63&DA/Me&0.2376198&0.03&0.1502&0.9098531$^{\rm f}$&(1)\\ 
		SDSS J2320&350.01674&+27.10662&31890&7.71&0.52&DA/Me&0.794531&1.0&0.0239&$-$&(4)\\ 
		SDSS J2343&355.80397&+15.68515&26801&7.84&0.56&DA/Me&0.5687399&1.06&0.0192&6.64$^{\rm b}$&(2)\\ 
		SDSS J0950&147.68317&+39.26165&41462&7.57&0.49&DA/Me&1.167341&1.5&0.0247&$-$&(10)\\ 
		SDSS J1317&199.46524&+67.53313&79476&6.99&0.44&DA/Me&3.38136&20.0&0.0365& 3.3820$^{\rm a}$&(2)\\ 
		SDSS J0803&120.76923&+12.30289&15964&9.0&1.20&DA/Me&0.2861439&0.25&0.0335& 0.5723126$^{\rm g}$&(2)\\
		SDSS J0858&134.57486&+39.16987&$-$&$-$&$-$&WD/Me&0.4268106&0.41&0.0410& 0.853573$^{\rm h}$&\\
		SDSS J1006&151.69061&+0.53463&26104&7.21&0.39&DA/Me&0.1055909&0.026&0.1077&0.2118$^{\rm g}$ &(1)\\
		SDSS J1016&154.08253&$-$2.04946&75536&8.40&0.91&DA/Me&0.3025928&0.21&0.0363&0.605190$^{\rm h}$&(1) \\
		SDSS J1145&176.29077&+38.22479&20000&9.0&1.20&DA/Me&0.09501896&0.026&0.0418&0.19003799$^{\rm g}$& (9)\\ \hline
		SDSS J0110&17.53785&+13.43789&25360&7.40&0.41&DA/Me&0.3326836&0.021&0.0239&0.3326867$^{\rm g}$& (11)\\
		SDSS J0314&48.71713&+2.10195&41140&8.02&0.69&DA/Me&0.3052987&0.062&0.3784&0.305297$^{\rm g}$&(3) \\
		SDSS J0836&129.07769&+43.44758&24726&7.64&0.46&DA/Me&0.1968954&0.13&0.0698&0.196898$^{\rm g}$& (2)\\
		LAMOST J090812&137.05016&+6.07252&17505&7.37&0.34&DA/Me&0.1494378&0.083&0.3273&0.149438$^{\rm d}$& (2)\\
		SDSS J0912&138.06824&+23.74517&30071&8.09&0.70&DA/Me&0.2635584&0.14&0.1225&0.2635582$^{\rm g}$&(2) \\
		SDSS J0939&144.95038&+32.96887&28389&7.75&0.52&DA/Me&0.3309915&0.3&0.0733& 0.33099$^{\rm d}$&(2)\\
		SDSS J0946&146.64359&+20.50089&10307&8.04&0.63&DA/Me&0.2528599&0.24&0.0925&0.25286122$^{\rm g}$&(2) \\
		SDSS J0957&149.33022&+23.71125&25891&7.55&0.43&DA/Me&0.15087312&0.063&0.1257&0.1508707$^{\rm g}$&(2) \\
		SDSS J1000&150.06317&+30.72522&14560&8.50&0.93&DA/Me&0.4515002&0.14&0.1514&0.4514995$^{\rm g}$&(2) \\
		SDSS J1013&153.48454&+27.40292&13000&9.0&1.15&DA/Me&0.12904056&0.02&0.1514& 0.12904$^{\rm d}$&(6)\\		
		SDSS J1106&166.61528&$-$1.08741&30071&7.46&0.42&DA/Me&0.4134605&0.53&0.0140&0.413462$^{\rm g}$&(2) \\
		SDSS J1108&167.02231&+65.36985&21045&7.53&0.41&DA/Me&0.3208369&0.35&0.1063&0.3208343$^{\rm g}$&(2) \\
		SDSS J1157&179.43685&+48.93838&11567&8.00&$-$&DA/Me&2.29114&22.0&0.0438& 2.2911685$^{\rm g}$&(1)\\
		SDSS J1226&186.62876&+30.64795&30071&7.41&0.400&DA/Me&0.2586904&0.23&0.1107&0.258687$^{\rm g}$&(2) \\
		SDSS J1239&189.76222&+65.82626&21470&7.76&0.50&DA/Me&0.661913&1.2&0.0528&0.6619085$^{\rm g}$ &(11)\\
		LAMOST J1249&192.49901&+3.95736&53842&7.28&0.42&DA/Me&0.743405&1.5&0.0214&0.743407$^{\rm i}$& (10)\\
		SDSS J1348&207.17369&+18.56961&16000&8.0&0.62&DA/Me&0.2484319&0.18&0.1063&0.24843148$^{\rm g}$&(6) \\
		SDSS J1408&212.19603&+29.84608&30000&7.75&0.52&DA/Me&0.19179001&0.09&0.0944&0.1917903$^{\rm g}$ &(6)\\
		SDSS J1411&212.89464&+10.4778&30419&7.74&0.52&DA/Me&0.1675101&0.06&0.1763&0.1675099$^{\rm g}$&(2)\\
		SDSS J1415&213.90176&+1.28836&73816&8.43&0.96&DA/Me&0.3443278&0.44&0.0672&0.34433084$^{\rm g}$ &(2)\\
		SDSS J1435&218.94948&+37.56086&12392&7.61&0.41&DA/Me&0.12563074&0.043&0.0844&0.1256311$^{\rm g}$&(2) \\
		LAMOST J1439&219.94843&$-$1.10189&76410&7.97&0.73&DA/Me&1.522581&8.2&0.0212& 1.522608$^{\rm e}$&(2)\\
		SDSS J1456&224.14262&+16.19379&20000&7.50&0.44&DA/Me&0.2291217&0.14&0.0180&0.2291202$^{\rm g}$&(6) \\
		SDSS J1519&229.77486&+50.11747&31270&7.66&0.49&DA/Me&0.3023353&0.24&0.0444& 0.30236455$^{\rm g}$&(11)\\
		SDSS J1539&234.90879&+27.10162&36572&7.31&0.40&DA/Me&0.2385547&0.05&0.0438& 0.238553$^{\rm g}$&(2)\\
		SDSS J1559&239.76931&+3.93978&48770&7.98&0.68&DA/Me&0.09434753&0.012&0.0363& 0.0943473$^{\rm g}$&(2)\\
	        SDSS J1620&245.12053&+63.07967&23551&7.12&0.31&DA/Me&0.2994283&0.32&0.482&0.299429$^{\rm j}$&(7)\\ 
		LAMOST J1724&261.02585&+56.33405&37999&7.15&0.37&DA/Me&0.3330190&0.13&0.1039&0.333028$^{\rm h}$&(13) \\
		SDSS J1730&262.51038&+33.56718&49042&7.345&0.42&DA/Me&0.1569469&0.086&0.1571&0.1569473$^{\rm g}$&(12) \\
		SDSS J2123&320.83431&+5.71489&21045&7.82&0.54&DA/Me&0.26204957&0.071&0.1215&0.2620477$^{\rm g}$&(1) \\
		\hline
	\end{tabular}
\end{table*}

\begin{table*}
	\centering
	\caption{The $\chi^2$-values derived for  different periods indicated by different peaks in periodograms for the newly discovered or period-improved PCEBs. The period indicated by the main peak is highlighted as bold for each PCEB.}
	\label{tab:chi2_table}
	\begin{tabular}{lccccccccr} 
		\hline
		Stars & frequencies & Periods & Amplitude&$\chi^{2}$ &Stars & frequencies & Periods & Amplitude&$\chi^{2}$\\
		  & (${\rm d^{-1}}$)  &(days) &(mag) &  && (${\rm d^{-1}}$)  &(days) &(mag) &\\
		\hline
		    & 8.2201419&{\bf 0.12165254} &0.0479&1.7488  &SDSS J1145 & 10.52421538 &{\bf 0.09501896} &0.0418& 7.5294 \\
SDSS J0029& 9.2228990&0.1084257 &0.0382& 1.8842 &                     & 12.52693508 & 0.07982799 &0.0336& 9.4481 \\
		    & 7.2173656 &0.1385524&0.0377 &2.95117&                     &                   &                    &                   &              \\ \hline
SDSS J0032& 6.4937663 &{\bf 0.1539945}& 0.0296&0.9183  &SDSS J1317& 0.295739       & {\bf 3.38136} &0.0365  & 2.7320\\
		    & 5.4910338 & 0.1821151&0.0293 & 1.9262 &                     &0.7042455  &  1.419959 &0.0309& 3.3896\\ \hline
SDSS J0253& 2.3450291&{\bf 0.4264303}&0.0374 & 0.9480 &SDSS J1424& 2.8170022    &{\bf  0.3549873} &0.0141   & 2.8010\\
		    & 1.3423156& 0.744981 &0.0343& 1.0685 &                      &3.8169625 &   0.2619884 &0.0138 & 4.3389\\ \hline
SDSS J0306& 1.8478892& {\bf 0.541158} &0.0217& 4.8067 &SDSS J1541 & 8.7699835    &{\bf  0.1140253}  &0.0599   & 2.2010\\
		    &2.8505955&0.3508039 &0.0194&4.8270 &                      &9.7699776     &   0.1023554 &0.0384 & 6.6393\\ \hline
SDSS J0747& 1.7297794& {\bf 0.5781084}&0.0395& 38.9850&LAMOST J1621& 0.312604  & {\bf 3.198935} &0.0080   & 2.9878\\
		   & 0.2728867 & 3.66453  &0.0321& 48.0663&                    &0.6901616   & 1.448936 &0.0072  & 3.1978\\ \hline
SDSS J0803 & 3.4947451&{\bf 0.2861439} &0.0335& 3.0446&SDSS J1638&2.2018283& {\bf 0.454168} &0.0230      & 1.4888\\
		    & 4.4975074 & 0.2223454&0.0322 & 4.6632&                        &3.2045401&  0.3120573 &0.0193   & 2.3150\\ \hline
SDSS J0858 & 2.3429596 &{\bf  0.4268106} &0.0410& 3.5121&SDSS J1705& 2.9364686&{\bf  0.3405451} &0.0987   & 1.7059\\
		    & 4.3457009 & 0.2301125&0.03333& 6.7661&                           &3.9391781& 0.2538601 &0.0847 & 1.7308\\ \hline		  
SDSS J0908& 7.2870098&{\bf 0.1372305}&0.1974  & 11.5378&SDSS J2130   & 3.8151455& {\bf 0.2621132} &0.1112& 3.8770\\
		   & 6.2843572& 0.1591253&0.1025& 46.4421&                         &2.8123848&  0.3555701 &0.0952& 7.2091\\ \hline
SDSS J0927 &3.2935787&{\bf 0.3036211}& 0.0190&1.7210&SDSS J2208     & 1.5284864& {\bf 0.654242 } &0.0173 & 3.7707\\ 
		    & 5.2962706& 0.1888121&0.0177 &2.2048&                         &0.525674&  1.90232 &0.0170& 4.2671\\ \hline
SDSS J0950& 0.8566477&{\bf 1.167341}& 0.0247&1.8047&SDSS J2302   & 4.2084351& {\bf 0.2376198}  &0.1502& 84.0070\\ 
		   &2.8594254& 0.3497206&0.0212& 1.9146&                        &3.2057117& 0.3119432  &0.1417& 245.3586\\	\hline
SDSS J1006 & 9.47051308 &{\bf 0.1055909} &0.1077& 4.1035&SDSS J2320& 1.258604&{\bf 0.794531} &0.0239& 2.2884\\
		  & 7.4677665 & 0.1339088&0.1055& 9.9490&                            &0.255855&  3.90845  &0.0170& 2.3862\\ \hline
SDSS J1016 & 3.3047713 &{\bf 0.3025928} &0.0363& 1.9146&SDSS J2343& 3.3526&{\bf 0.5687399} &0.0192& 1.3381\\
		  & 4.3075007 & 0.2321532&0.0341& 3.2178&                            &0.7555755&  1.323494 &0.0158& 1.9402\\ \hline
SDSS J1038 & 1.1975403 &{\bf 0.835045} & 0.0125& 1.2491&&& && \\ 
		  &3.2002366& 0.3124769&0.0116& 1.7343&&& &&  \\ \hline
	\end{tabular}
\end{table*}

\begin{table}
	\centering
	\caption{The photometric data for SDSS J1541.}
	\label{tab:SDSS1541_table}
	\begin{tabular}{lccr} 
		\hline
		Bands & Wave length & Magnitude & Flux \\
		 &($\micron$)  & (mag)    &($\mu$Jy)   \\
		\hline
		$u'$ & 0.3572 & 16.830(7) &699.8(4.5)\\
		$g'$ & 0.4640 & 16.809(4) & 684.7(2.5)\\
		$r'$ & 0.6185 &17.174(5) & 488.8(22.0)\\
		$i'$ & 0.7439&17.398(7)   & 393.0(2.5)\\
		$z'$ & 0.8897 & 17.533(17) & 343.4(5.3)\\
		$g$ & 0.4810& 16.886(5) & 639.1(2.9)\\
		$r$ & 0.6155&17.250(8)   & 457.1(3.4)\\
		$i$ & 0.7503& 17.579(6) & 337.6(1.9)\\
		$z$ & 0.8668 & 17.742(20)& 290.5(5.3)\\
		$y$ & 0.9613&17.176(28)   &297.6(7.5) \\
		$G$ & 0.5858 & 17.032(4) & 501.0(1.8)\\
		$G_{\rm bp}$ & 0.5044 & 16.967(8)& 588.0(4.3)\\
		$G_{\rm rp}$ & 0.7692&17.164(15)   & 344.8(4.7)\\
		$W_1$ & 3.3526& 16.428(67)& 276.1(16.5)\\
		$W_2$ &4.6028& 16.299(205)& 192.9(33.1)\\
		\hline
	\end{tabular}
\end{table}

\subsubsection{SDSS J0253}
SDSS J025301.60-013006.96 (SDSS J0253) was first identified as a DA+M binary based on the spectra from SDSS DR9 by \citet{Li2014}. The atmospheric and physical
parameters for this object were derived based on the analysis of its spectra from LAMOST and SDSS. It is found that this binary system contains a
hot WD and an M-type dwarf \citep{Li2014,Guo2015,Kepler2015,Ren2018}.

It is found in Fig.~\ref{fig:spec_figure} that this object shows the convincing emissions at H$\alpha,\beta$ and Ca II triplet which might  arise from an irradiated component in this object, implying that SDSS J0253 might be a short-period PCEB. Therefore, it is possible that the variability in its luminosity can be detected based on the photometric observations if its orbital inclination is appropriate. In addition, \citet{Holl2023} had found a spurious signal with a frequency of 1.37165769 ${\rm d^{-1}}$ (corresponds to a period of 0.7290448 days) probably related to time-dependent scan angle for this object based on Gaia DR3 $G$-band time series data. In order to determine whether this spurious period is its orbital period or not, we collect the photometric data from ZTF DR19 to study the variability in the luminosity of SDSS J0253. As a result, the total of 464 data points in
$g$-band, 476 data points in $r$-band and 46 data points in $i$-band are obtained. Then the light curves are analyzed after some data points with a large scatter are ignored,
its orbital period is derived to be 0.4264210(19) days in $g$-band and  0.4264303(15) days in $r$-band, respectively and the results are also listed in Table~\ref{tab:lc_table}, and only one periodogram indicating its orbital period based on $r$-band data is shown in Fig.~\ref{fig:ZTF00321_figure} because of the same reason as that for SDSS J0032. Although there is a peak with a similar power to the main one and a frequency of 1.3423156 ${\rm d}^{-1}$ [corresponds to a period of 0.7449813(89) days] based on Fig.~\ref{fig:ZTF00321_figure}, which is still different from the signal frequency obtained by \citet{Holl2023}, so the period of the signal discovered by \citet{Holl2023} might be a spurious signal related to time-dependent scan angle rather than the orbital period of this object. Meanwhile, it is found in Table~\ref{tab:chi2_table} that the $\chi^2$-value based the period indicated by the main peak is also smaller than that derived from this peak, implying that the period indicated by the main peak is the orbital period of this object. The phase-folded light curves in $g$, $r$ and $i$-bands are drawn in Fig.~\ref{fig:ZTF0032_figure}(b), and the variability in its luminosity is caused by the reflection effect or star spot. 

\subsubsection{SDSS J0306}

SDSS J030607.19-003114.44 (SDSS J0306, also named KUV03036 -0043) was first observed spectroscopically by KISO Schmidt ultraviolet
excess survey and then was classified as a DA+dM binary  \citep{Wegner1987}. Subsequently the different atmospheric and physical
 parameters for the DA star in this object had been obtained by the various investigators, it was found that this WD might be a He-WD
 \citep{Schreiber2010,Debes2011} or a C/O-WD \citep{Tremblay2011,Limoges2010,
Morgan2012,Rebassa2010,Gianninas2011}.
Although KUV 03036-0043 was identified as a WD+M4/M5 binary by \citet{Raymond2003} and \citet{Kleinman2013},
however it was argued as a close binary by \citet{Silvestri2007} and \citet{Morgan2012}  who estimated an upper limit of 2.66 days
for the orbital period of this object based on several RV measurements.

The radial velocities were determined for the DA WD component by some previous investigators \cite[][]{Schreiber2010,Rebassa2010,Morgan2012,Dietz2020}. It was found that its radial velocity exhibits a large peak to peak variation \citep[about 273.30 \kms,][]{Schreiber2010}. Meanwhile, Balmer emission lines are evidently presented in its optical spectra (see Fig.~\ref{fig:spec_figure}), they might be caused by magnetic activity or irradiation of M-type dwarf due to its hot WD companion. These observational characteristics suggest that KUV 03036-0043 might be a short-period PCEB which provides a favorable opportunity for researchers to observe the variability in its luminosity. So we collect the photometric data from ZTF DR19 to investigate the variability in the luminosity of SDSS J0306. As a result, the total 422 data points in $g$-band and 428 data points in $r$-band are
obtained. After some data points with a large scatter are neglected, the light curves are analyzed. As the objects mentioned above, only one periodogram indicating its orbital period based on $r$-band data is also shown in Fig.~\ref{fig:ZTF00321_figure}. Meanwhile, the $\chi^2$-values are obtained for its $r$-band light curves based on two orbital periods indicated by the main peak and another peak with a similar power to the main one and a frequency of 2.8505955 ${\rm d}^{-1}$ ($P_{\rm orb}=0.3508039$ days) and they are also listed in Table~\ref{tab:chi2_table}. It is found in Table~\ref{tab:chi2_table} that the $\chi^2$-value based on the orbital period indicated by the main peak is also a smaller one, suggesting that the period implied by the main peak should be its orbital period, then its orbital period is derived to be 0.541179(65) days in $g$-band and 0.541158(54) days in $r$-band, respectively. Its phase-fold light curves are displayed in Fig.~\ref{fig:ZTF0032_figure}(c).  As seen from Fig.~\ref{fig:ZTF0032_figure}(c), this object is a short-period PCEB with an orbital period of about 12.9878 hours, and its luminosity change might be caused by  star spot due to magnetic activity or reflection effect. In addition, the orbital period obtained for this object by us is indeed smaller than its upper limit listed in \citet{Morgan2012}, suggesting that the orbital period obtained for this object might be correct.

\subsubsection{SDSS J0747}
SDSS J074730.57+430403.65 (SDSS J0747) was first identified as WD+dM binary according to its spectrum from SDSS by \citet{Raymond2003}. The atmospheric and
physical parameters for this object were obtained through a analysis of the spectrum from SDSS \citep{Schreiber2010,Rebassa2010}. It was found that this object is a DA+M4/M5
binary \citep{West2008,Kleinman2013}. The radial velocities for SDSS J0747 were analyzed by  \citet{Rebassa2010} and \citet{Schreiber2010} who gave
a peak to peak radial velocity variation of 328.90 \kms. This suggests that SDSS J0747 might be a short-period PCEB. Meanwhile, Balmer emission lines evidently
present in its optical spectra from SDSS, this also indicates that SDSS J0747 might be a short-period PCEB and they might arise due to magnetic activity or irradiation of
the cool component by a hot WD  \citep[more than 15,000 K,][]{Rebassa2010,Schreiber2010}.

The radial velocities were derived to be -487.6(248.4) and 177.0(417.7) \kms for M-type dwarf and WD components respectively by \citet{Morgan2012}  who also estimated an upper limit of 1.33 days for the orbital period of this object, suggesting that the variability in its luminosity might be detected. In order to find the variability in its luminosity, we collect the photometric data for it from ZTF DR19, the total 570 data points in $g$-band and 2394 data points in $r$-band are obtained, the light curves are also analyzed after some data points with a large scatter are removed, and the periodogram manifesting the orbital period for
SDSS J0747 is also displayed in Fig.~\ref{fig:ZTF00321_figure}.  The $\chi^2$-values derived for the $r$-band light curves based on the orbital periods indicated by the main peak and another peak with a frequency of 0.2728867 ${\rm d}^{-1}$ ($P_{\rm orb}=3.66453$ days) are also listed in Table~\ref{tab:chi2_table}. As seen from Table~\ref{tab:chi2_table}, the $\chi^2$-value based on the main peak is a smaller one. This suggests that the period indicated by the main peak should be the orbital period of this object. Its orbital period is thus derived to be 0.5781084(17) days in $r$-band and the result is listed in Table~\ref{tab:lc_table} .  The phase-folded light curves of SDSS J0747 are plotted in Fig.~\ref{fig:ZTF0036_figure}(a). It is found in Fig.~\ref{fig:ZTF0036_figure}(a) that this object should be an eclipsing PCEB with a short orbital period of about 13.875 hours.  In addition, the orbital period determined for
this object is indeed smaller than its upper limit listed in \citet{Morgan2012}, also implying that the orbital period derived for this object might be correct.

\begin{figure*}
	\includegraphics[width=\columnwidth]{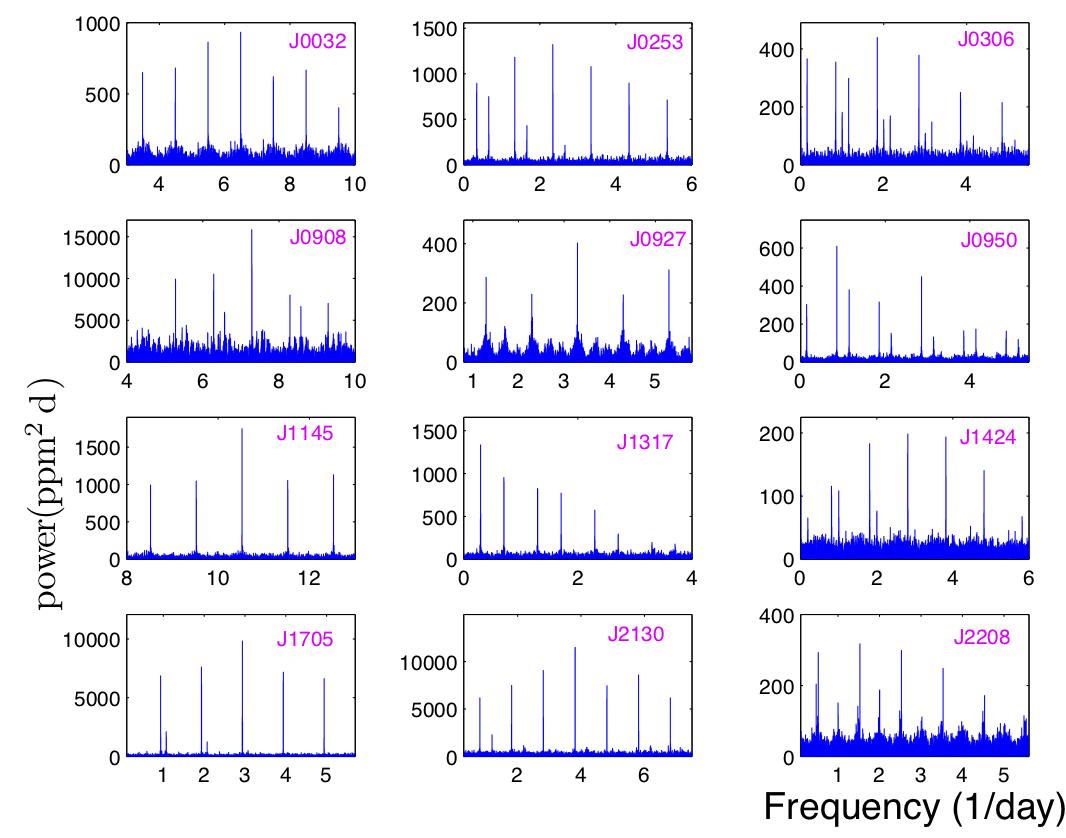}
	\includegraphics[width=\columnwidth]{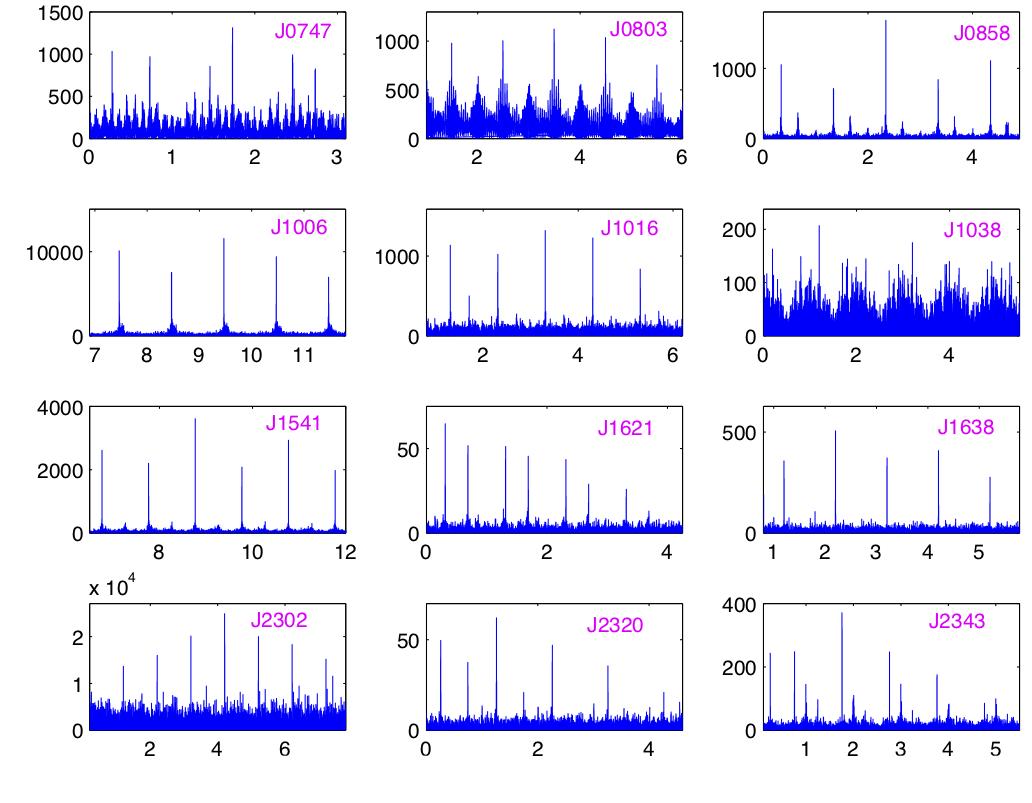}
    \caption{The periodograms indicating orbital periods derived from $r$-band ZTF photometric observations for newly discovered or period-corrected PCEBs (only SDSS J2320 based on $g$-band observations).}
    \label{fig:ZTF00321_figure}
\end{figure*}

\begin{figure*}

	\includegraphics[width=8.5cm, height=8.5cm]{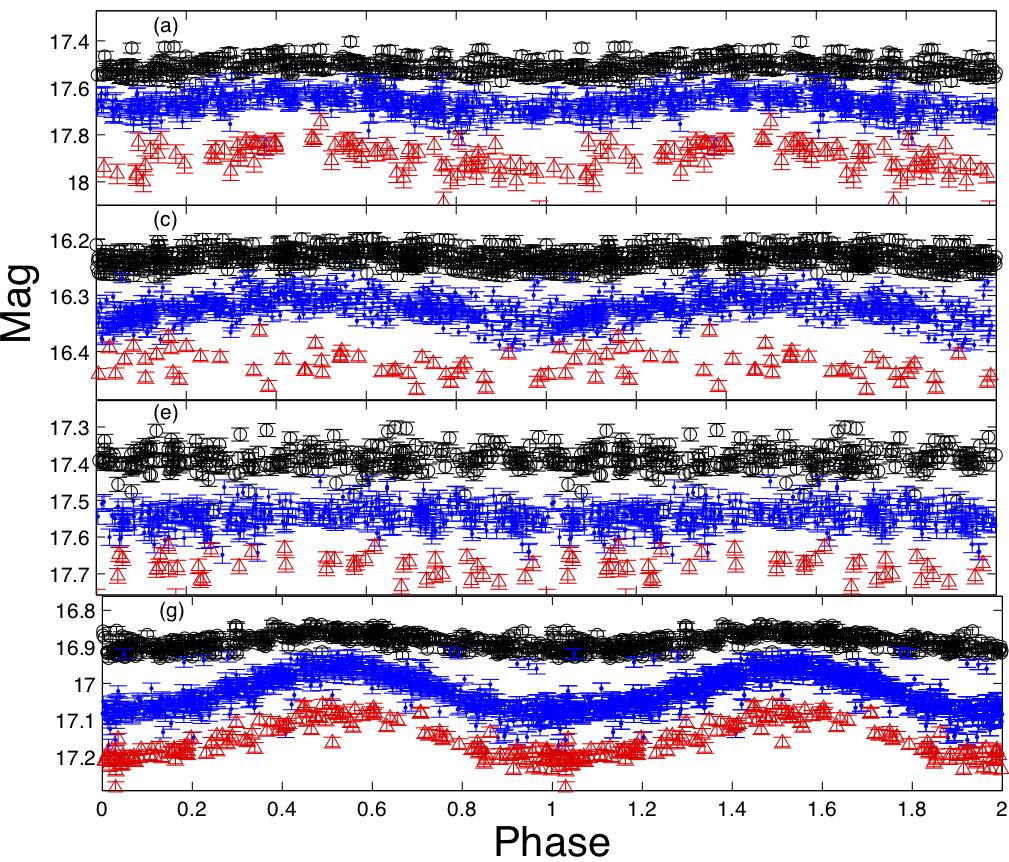}
	\includegraphics[width=8.5cm, height=8.5cm]{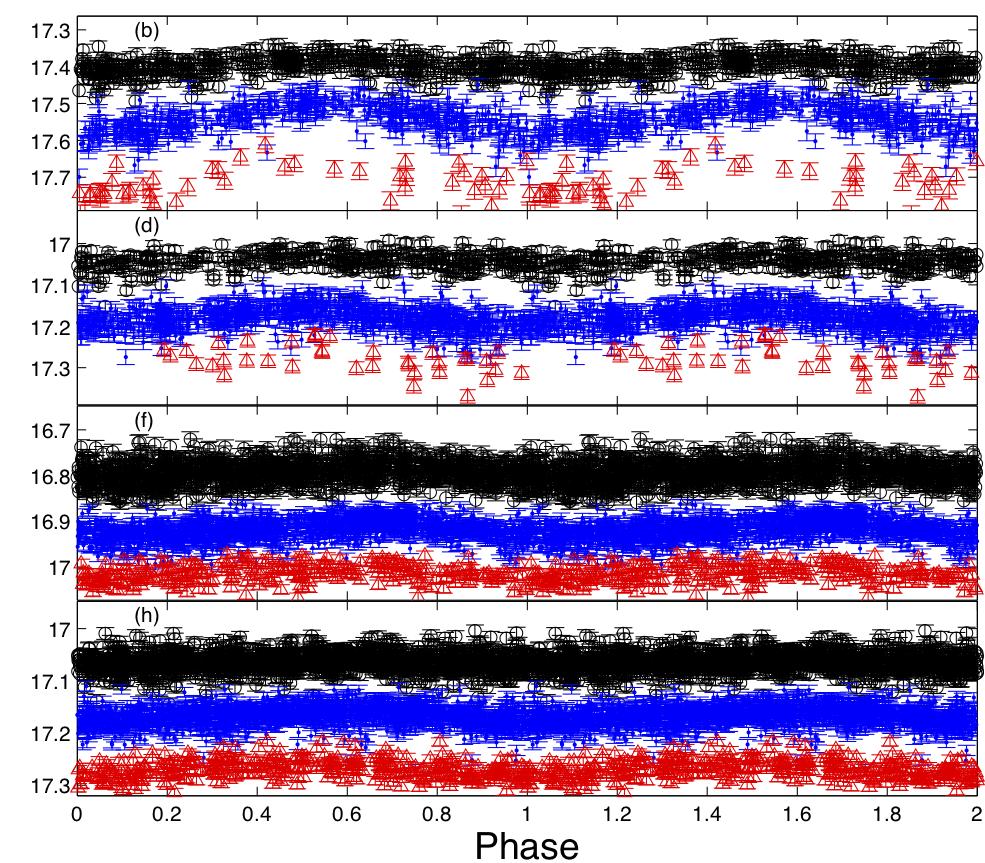}
    \caption{The light curves in $g$, $r$ and $i$-bands for 8 PCEBs only with hydrogen emission line(s). In each panel, black open dots represent the $g$-band observations, the blue solid dots indicate the $r$-band  observations, and the red  open triangles represent the $i$-band ones.}
    \label{fig:ZTF0032_figure}
\end{figure*}

\subsubsection{SDSS J0908}

SDSS J090847.38+613141.43 (hereafter SDSS J0908) was first identified as a WDMS by \citet{Morgan2012} , then the atmospheric and physical parameters for
the DA star in this binary system were given as the followings:  $T_{\rm WD}=11,000$K, ${\rm log}g=8.25$, $M_{\rm WD}=0.827M_{\sun}$, $V_{\rm WD}=37.6$ \kms and
$V_{\rm dM}=-124.6$ \kms. Another estimation for the radial velocity of DA WD in this object was $-40.436\pm28.598$ \kms \citep{Tsantaki2022}. These works indicate that this object might be a short-period PCEB. 

 An upper limit on the orbital period for this object was estimated to be 137.43 days by \citet{Morgan2012}. However, as seen Fig.~\ref{fig:spec_figure},  the Balmer emission lines also arise in the optical spectrum from SDSS, this suggests that this object might
be a short-period PCEB. In order to obtain the accurate period for it, we collect the photometric data from ZTF DR19 for SDSS J0908, then the total 649 data points in $g$-band and 780 data points in $r$-band are
obtained. The light curves of SDSS J0908 are analyzed through the same method as that mentioned above, and only one periodogram indicating the
orbital period for SDSS J0908 based on $r$-band is also shown in Fig.~\ref{fig:ZTF00321_figure}. As seen form Fig.~\ref{fig:ZTF00321_figure}, it is difficult to rule out some possible orbital periods indicated by other peaks with a similar power to the main peak. In order to find the orbital period for this object, we calculate the $\chi^2$-values for the $r$-band light curves based on the periods indicated by the main peak and another peak with a frequency of 6.2843572 ${\rm d}^{-1}$ ($P$=0.1591253 days), which are listed in Table~\ref{tab:chi2_table}. It is found in Table~\ref{tab:chi2_table} that the $\chi^2$-value based on the main peak is also a smaller one, implying that the period indicated by the main peak should be the orbital period of this object, then its orbital period is derived
to be  0.1372302(10) days in $g$-band, 0.1372305(7) days in $r$-band, respectively and the results are listed in Table~\ref{tab:lc_table}.  Its phase-folded light curves for SDSS J0908 are shown in Fig.~\ref{fig:ZTF0036_figure}(b). It is found in Fig.~\ref{fig:ZTF0036_figure}(b) that SDSS J0908 should be also a close eclipsing PCEB. Meanwhile, the orbital period derived for this object is much smaller than its upper limit estimated by \citet{Morgan2012}.

\subsubsection{SDSS J0927}

SDSS J092712.02+284629.28 (SDSS J0927) was first suspected to be a DA WD by \citet{Pesch1983}, then
it was identified as a DA WD by \citet{Wagner1988}. Thereafter, it was reclassified as a DA/M binary \citep[e.g.][]{Kleinman2013,Guo2015}.
The atmospheric and physical parameters for both components of this object were derived based on the the analysis of the spectra from
LAMOST or SDSS \citep{Rebassa2010,Gianninas2011,Morgan2012,Guo2015}. The atmospheric and physical parameters were derived to
be $T_{\rm eff}=22,037$ K, ${\rm log}g=7.80$,  and $M_{\rm WD}=0.52M_{\sun}$ by \citet{Rebassa2010} who gave a distance of 237 pc away from the Earth, which is closest to a distance of about 235.4$\pm$5.5 pc indicated by its parallax from $Gaia$ DR3 \citep{Gaia2021}. The radial velocities of two components in SDSS J0927
were derived to be $-73.5$ and 133.1 \kms for the M-type star and DA star, respectively \citep{Morgan2012}  who estimated an upper limit of 13.46 days for the orbital period of this object, this suggests that this object might be a PCEB.

H$\alpha$ emission line is evidently presented in its optical spectra from SDSS and LAMOST (see Fig.~\ref{fig:spec_figure}), this also implies that this object might be a
short-period PCEB. We also collect the photometric data for SDSS J0927 from ZTF DR19. As a result, the total of 356 data points in $g$-band and 758 data points
in $r$-band are obtained. Then the light curves are also analyzed after some data points with a large scatter are neglected. Its orbital period is determined to
be 0.3036308(63) days in $g$-band and 0.3036211(47) days in $r$-band, respectively. The results are listed in Table~\ref{tab:lc_table} and only one periodogram based on
$r$-band data is shown in Fig.~\ref{fig:ZTF00321_figure}.  As the same as the objects mentioned above, we have to calculate the $\chi^2$-values for its $r$-band light curves based on the different orbital periods indicated by the main peak and another peak with a similar power to the main peak (see Table~\ref{tab:chi2_table}). The results suggest that the period indicated by the main peak should be the orbital period of this object, then the phase-folded light curves for SDSS J0927 are displayed in Fig.~\ref{fig:ZTF0032_figure}(d). As seen from Fig.~\ref{fig:ZTF0032_figure}(d), SDSS J0927 should be a PCEB with an orbital period of about 7.287 hours.
In addition, the orbital period obtained for this object is indeed smaller than its upper limit given by \citet{Morgan2012} and thus might be correct.

\begin{figure}
	\includegraphics[width=\columnwidth]{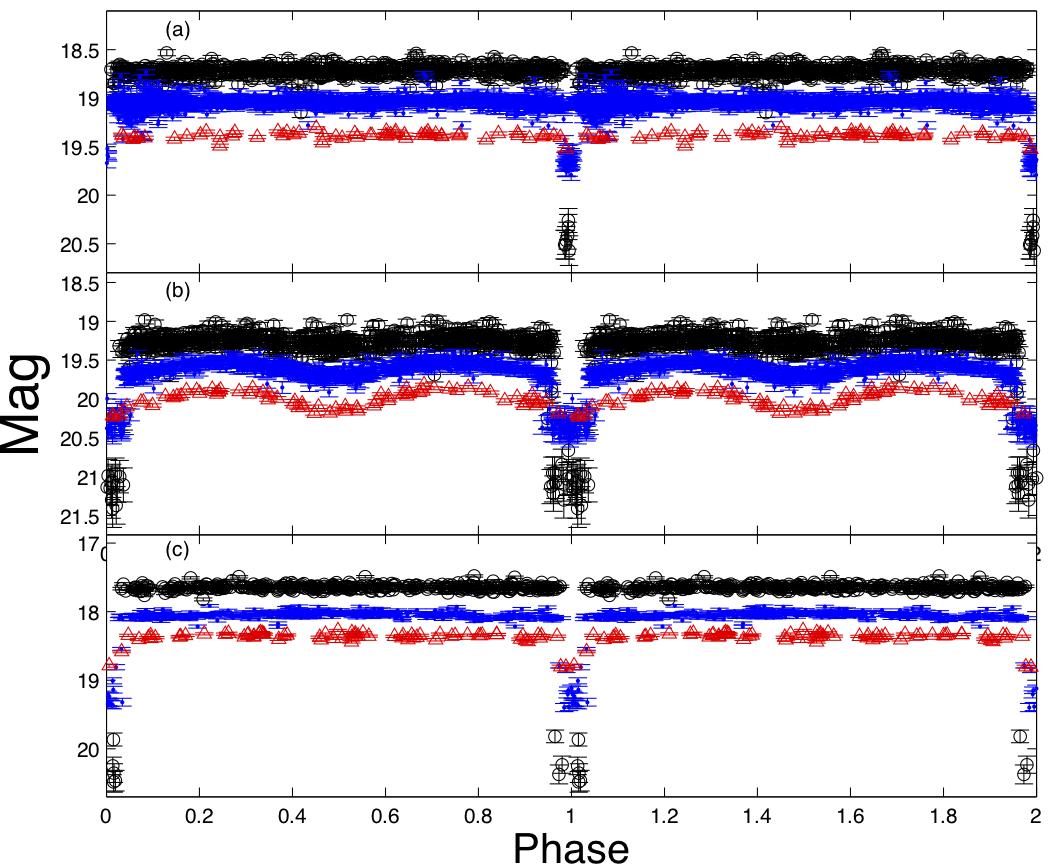}
    \caption{Same as Fig. 5, but for three eclipsing PCEBs (SDSS J0747, SDSS J0908 and SDSS J2302) .}
    \label{fig:ZTF0036_figure}
\end{figure}

\begin{figure}
	\includegraphics[width=8.5cm, height=7.5cm]{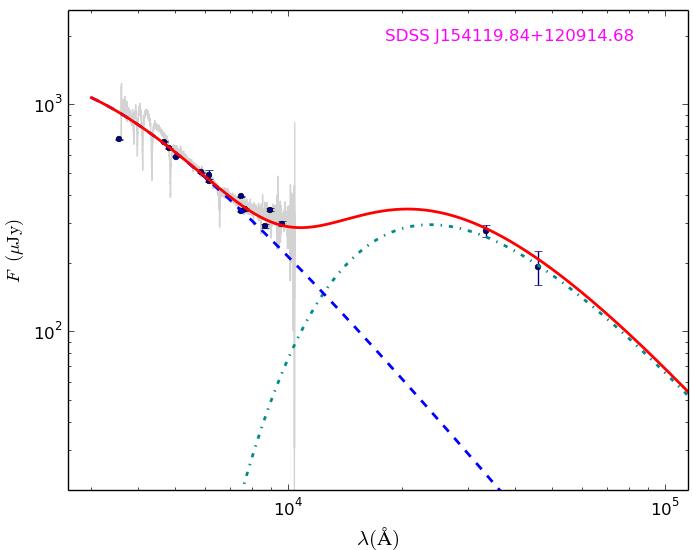}
    \caption{The fitting result of SED of SDSS J1541 based on the atmospheric parameters obtained in this work. Solid dots represent the observed fluxes derived from
    the photometric magnitudes, the dashed line (blue) indicates the contribution of the white dwarf, and the dot-dashed line (darkcyan)
    means the contribution of a cool dwarf. The solid line (red) represents the contribution of both white dwarf and M-type dwarf, the gray line denotes its observed spectrum
    from SDSS.}
    \label{fig:SED1541_figure}
\end{figure}

\subsubsection{SDSS J1038}
SDSS J103837.22+015058.48 (hereafter SDSS J1038) was classified as a close WD+MS binary by \citet{Silvestri2006} based on the spectra from
SDSS DR4. Its atmospheric and physical parameters had been obtained by other investigators based on the spectra from SDSS
\citep[e.g.][]{Rebassa2010,Debes2011,Tremblay2011,Morgan2012,Kleinman2013,Bedard2020}. Thereafter, these parameters were derived from the optical spectrum from LAMOST by \citet{Ren2018} who gave the similar results as those based on the spectra from SDSS. The radial velocities for both components of this object were derived to be
$-180.4$ and 157.4 \kms for the M-type star and WD respectively by \citet{Morgan2012}  who also estimated an upper limit of 2.68 days for the orbital period of this object based on several RV measurements, this implies  that this object might be a short-period PCEB. In order to find its accurate period, we collect the photometric data for this object from ZTF DR19 to investigate
the variability in its luminosity. The total of 233 data points in $g$-band and 345 data points in $r$-band are obtained, then the light curves of SDSS J1038 are analyzed if
some data points with a large scatter are not  taken into account and the periodogram implying the orbital period for this object is displayed in Fig.~\ref{fig:ZTF00321_figure}. As seen from Fig.~\ref{fig:ZTF00321_figure},  there are some peaks with a similar power to the main peak in its periodogram, we must calculate the $\chi^2$-values for its $r$-band light curves based on the periods indicated by the main peak and another peak with a frequency of 3.2002366 ${\rm d}^{-1}$ (corresponding to $P_{\rm orb}$=0.3124769 days), which are listed in Table~\ref{tab:chi2_table}. It is found in Table~\ref{tab:chi2_table} that the $\chi^2$-value based on the main peak is the smaller one, suggesting that the period indicated by the main peak should be the orbital period. At last, its orbital period is derived to be 0.835045(35) days according to the photometric data in $r$-band and
its phase-folded light curves are shown in Fig.~\ref{fig:ZTF0032_figure}(e).  As seen from Fig.~\ref{fig:ZTF0032_figure}(e), SDSS J1038 is a short-period PCEB although the amplitude of the variation in its luminosity is small because of a small orbital inclination.  Meanwhile, the orbital period obtained for this object by us is really smaller than its upper limit given by \citet{Morgan2012} and thus should be correct.

\subsubsection{SDSS J1424}
SDSS J142417.74+443225.00 (SDSS J1424) was  identified as a WD+MS binary by \citet{Heller2009} and \citet{Rebassa2010} because of its ultraviolet excess.
The effective temperature of the main sequence component in this object was derived to be 5116(72) K  \citep{Miller2015} or 5206 K \citep{Tonry2018}, which
corresponds to an effective temperature of a $\sim$K1-type dwarf \citep{Bell1989}. The radial velocities of this object were determined as 55.00 (Na I) or 36.50
(${\rm H}\alpha$) \kms based on SDSS spectrum \citep{Rebassa2010}. Another radial velocity for this object was derived to be 319.27 \kms based on Gaia
BP/RP spectrum \citep{Verberne2024}, suggesting that SDSS J1424 shows a large variability in its radial velocity and thus might be a short period PCEB.

Meanwhile, H$\alpha$ emission line is also evidently presented in its optical spectra (see Fig.~\ref{fig:spec_figure}), this also implies that this object might be a
short-period PCEB. In order to find its orbital period, we collect the photometric data for this object from ZTF DR19, and analyze its light curves in $g$, $r$ and
$i$-bands. An orbital period is derived to be 0.3549873(29) days for SDSS J1424 from its $r$-band photometric data and the periodogram implying its orbital period is
shown in Fig.~\ref{fig:ZTF00321_figure}. It is also difficult to rule out some possible periods indicated by the peaks with a similar power to the main one for this object based on Fig.\ref{fig:ZTF00321_figure}. By using the same method as that used for the objects mentioned above, it is found that the period indicated by the main peak should be the orbital period of this object (see Table~\ref{tab:chi2_table}). At last, its phase-folded light curves are shown in Fig.~\ref{fig:ZTF0032_figure}(f). It is found in Fig.~\ref{fig:ZTF0032_figure}(f) that SDSS J1424 is indeed a short period PCEB with an orbital period of about 8.5187 hours.

\subsubsection{SDSS J1541}

SDSS J154119.84+120914.68 (SDSS J1541) was first identified as a DA WD by \citet{Girven2011} who had given
the effective temperature [$T_{\rm eff}=19,000$ K] and surface gravity [${\rm log}g=8.50$]
based on a spectrum
obtained by SDSS on 2011 June 25. These spectral
parameters of SDSS J1541 were improved by
\citet{Kepler2015} and \citet{Kepler2019}, who gave the effective temperature [$T_{\rm eff}=25,464(129)\ {\rm K}$],
surface gravity [${\rm log}g=7.480(16)$], and mass [$M_{\rm WD}=0.454(6)\,{\rm M_{\rm \sun}}$] for
this object based on the same spectrum as that used in \citet{Girven2011}.
It is found in Fig.~\ref{fig:spec_figure} that the emission lines are evidently presented at H$\alpha$ and Ca II triplet, suggesting that
this object should be companied by a cool star. The spectral features of this object are almost dominated by its WD companion. Apart from emission characteristics of the
companion star, its other spectral features are difficultly found out from the spectrum of this object. Therefore, it is necessary to obtain the properties of its cool
companion by analyzing the spectral energy distribution (SED) of this object. However, there is an evident difference in its spectral parameters
provided by previous investigators, thus we have to verify its spectral parameters through spectral analysis since the result of SED analysis strongly depends on
the spectral and physical parameters of the WD component.
By using of a grid of WD model atmospheres \citep{Koester2010}, we analyzed its SDSS spectrum
again under an assumption that the WD has redshift owing to proper motion and strong
gravity of this object, we obtained the effective
temperature [$T_{\rm eff}=25,369(142)$ K], surface
gravity [${\rm log}g=7.44(2)$] and a radial velocity [$51.0{\rm(\pm4.0)\
km\ s^{-1}}$] for this object.
we derived the mass [$M_{\rm WD}=0.44(1) M_{\sun}$], cooling age [$\tau_{\rm c}=45.8\pm 3.9$ Myr], and radius [$R_{\rm WD}=0.0208(4)\ R_{\rm \sun}$] for this WD on the basis of
a recently updated version of the cooling models \citep{Bergeron1995}. At last, based on a well known
equation $d=\sqrt{\frac{\pi}{a}}\frac{R_{\rm WD}{\rm [R_{\sun}}]}{1{\rm pc}}$ \citep[where $a=F_{\rm obs, \lambda}/F_{\rm ast, \lambda}$ is a ratio of the observed spectral flux on the earth to the astrophysical flux
at the stellar surface,][]{Heller2009}, the spectroscopic distance of this WD
is estimated to be of 369($\pm$9) pc away from the Earth, which is consistent with a distance of 370.3$\pm$8.3 pc derived through its parallax \citep[$2.7003\pm0.0636$,][]{Gaia2021}. These results are in good agreement with those derived by \citet{Kepler2015,Kepler2019}.

Based on the ZTF photometric data, this object was discovered a variable star with an orbital period of 0.10232 days in $g$-band and 0.11403 days in $r$-band \citep{Chen2020}. This also suggests that this object should not be a single white dwarf, but a binary containing a hot DA star. Meanwhile, we fit its ${\rm H\alpha}$ emission line by
using a Gaussian profile in detail, the radial velocity of its cool component is derived to be $-183.0\pm1.3$ \kms, which is much higher than that of WD \citep[+51(4)\kms in this work,
and 19(4)\kms in][]{Kepler2019}. Therefore, the mass of its companion should be much lower than that of the DA star. 

In order to obtain the properties of the cool component of SDSS J1541, it is necessary to investigate the SED of
SDSS J1541 based on the photometric magnitudes from optical to infrared bands. The optical photometric magnitudes are obtained from  SDSS  DR7 \citep[$u'g'r'i'z'$,][]{Abazajian2009}, Gaia DR3 \citep[$GG_{\rm bp}G_{\rm rp}$][]{Gaia2021} and Pan-StaRRs DR1  \citep[$grizy$,][]{Chambers2016}.
The near-infrared (near-IR) photometry data are taken from WISE \citep[$W_1$ and $W_2$,][]{Wright2010}.
The magnitude and flux density in each passband for
SDSS J1541 are listed in Table~\ref{tab:SDSS1541_table}, and they are plotted in Fig.~\ref{fig:SED1541_figure}
with solid dots. As seen from
Fig.~\ref{fig:SED1541_figure}, SDSS J1541 indeed shows IR excesses from $z$-band to $W_2$,
and thus this hot WD should be companied a cool companion.

IR excesses of WDs are usually explained by the existence of a debris disk or
a cool companion (a cool dwarf or even a planet). The parameters of the cool component
of WDs are usually derived based on
the SED analysis through
a least $\chi^{2}$ methods \citep[][and references therein]{Girven2011}.  Using this method, the SED of SDSS J1541 is analyzed, then
the temperature and radius for its cool component are derived to be 2,018(207) K and 0.438$R_{\sun}$, corresponding to a late M-type star with a
mass of about 0.080$M_{\sun}$ according to an effective-mass relationship with an age of 5 Gyr \citep{Baraffe2003,Rebassa2007}. Therefore SDSS J1541 is a short-period
PCEB composed of a hot WD and a late M-type dwarf. Although the orbital period for this object has been derived by \citet{Chen2020}, however there is a
difference (about 17 min) between a period determined by $g$-band data and another one based on $r$-band data. We collect the photometric data
from ZTF DR19, and obtain 375 data points in $g$-band, 605 data points in $r$-band and 133 data points in $i$-band, respectively. Then the light curves are
analyzed through a method used in SDSS J0029, and only one periodogram implying its orbital period based on $r$-band data is shown in Fig.~\ref{fig:ZTF00321_figure}.  Its orbital period is derived to be 0.1140249(11) days in $g$-band, 0.1140253(10) days in $r$-band and 0.1140261(11) in $i$-band, respectively, and the results are also listed in Table~\ref{tab:lc_table}. The light curves based on an orbital period determined by $r$-band data for SDSS J1541 are displayed in Fig.~\ref{fig:ZTF0032_figure}(g), It is found in
Fig.~\ref{fig:ZTF0032_figure}(g) that SDSS J1541 is a PCEB with an orbital period of about 2.7366 hrs and the variation in its luminosity might be caused by
reflection effect or a dark spot due to stellar activity. Our results on the orbital period of this object are in well agreement with that derived by \citet{Chen2020} based
on $r$-band data, and a different orbital period derived for this object by them based on $g$-band data might be caused by an selection of wrong peak due to scattered data points or fewer data points in $g$-band at that time. In fact, it is found in Fig.~\ref{fig:ZTF00321_figure} that a signal with a similar power to main peak and a frequency of 9.769997632 ${\rm d}^{-1}$, corresponds to an orbital period of 0.102355417 days, which is in agreement with that given by \citet{Chen2020} based on $g$-band data. This suggests that the scattered data points or the number of all data points might lead to a wrong peak selection. Meanwhile, it is found in Table~\ref{tab:chi2_table} that $\chi^2$-value according to the period indicated by the main peak is smaller than that based this peak, suggesting that the orbital period should be the period indicated by the main peak as the objects mentioned above.

\subsubsection{LAMOST J1621}
LAMOST J162112.62+411809.81 (also named KUV 16195+4125) was first observed spectroscopically by the KISO Schmidt ultraviolet excess survey
and discovered as a DA+dM binary \citep{Wegner1990}. After that, it was found that this binary system can be resolved with Hubble Space telescope \citep{Farihi2006}. The atmospheric and physical parameters for the DA star were derived to be $T_{\rm eff}=14,090(457)$ K,
${\rm log}g=7.93(6)$ and $M_{\rm WD}=0.57_{\sun}$ \citep{Limoges2010}. The similar results had been obtained again by other investigators based on the spectra from
KISO or LAMOST \citep{Gianninas2011,Guo2015,Ren2018}. 

Meanwhile, it is found in Fig.~\ref{fig:spec_figure} that the emission feature at H$\alpha$ also can be seen from its LAMOST spectrum clearly, implying that this object might be a close PCEB and thus the variability in its luminosity might be detected through photometric observations. We collect the photometric data from ZTF DR19 for it, the total of 1330 data points
in $g$-band, 1321 data points in $r$-band and 379 data points in $i$-band are obtained. The light curves are analyzed and only one periodogram indicating its orbital period based on $r$-band data is displayed in Fig.~\ref{fig:ZTF00321_figure} and the results are listed in Table~\ref{tab:lc_table}. In addition, it is found in Table~\ref{tab:chi2_table} that the $\chi^2$-value derived for its $r$-band light curve based on the period indicated by the main peak is smaller than that based on a peak with a similar power to the main one and a frequency of 0.6901616 ${\rm d}^{-1}$ ( corresponding to a period $P=1.448936$ days), so that the period indicated by the main peak should be its orbital period, then the orbital period is derived to be 3.198935(91) days in $r$-band and 3.200093(145)
days in $i$-band, respectively. then the light curves based on the orbital period determined by photometric observations in $r$-band are displayed in  Fig.~\ref{fig:ZTF0032_figure}(h).
It is found in Fig.~\ref{fig:ZTF0032_figure}(h) that LAMOST J1621 indeed exhibits the periodic variation in its luminosity with a short period and is thus a PCEB with a short orbital period. The variability in the luminosity of LAMOST J1621 might be a result of reflection effect or a star spot owing to magnetic activity.

\begin{figure*}
	\includegraphics[width=8.5cm, height=8.5cm]{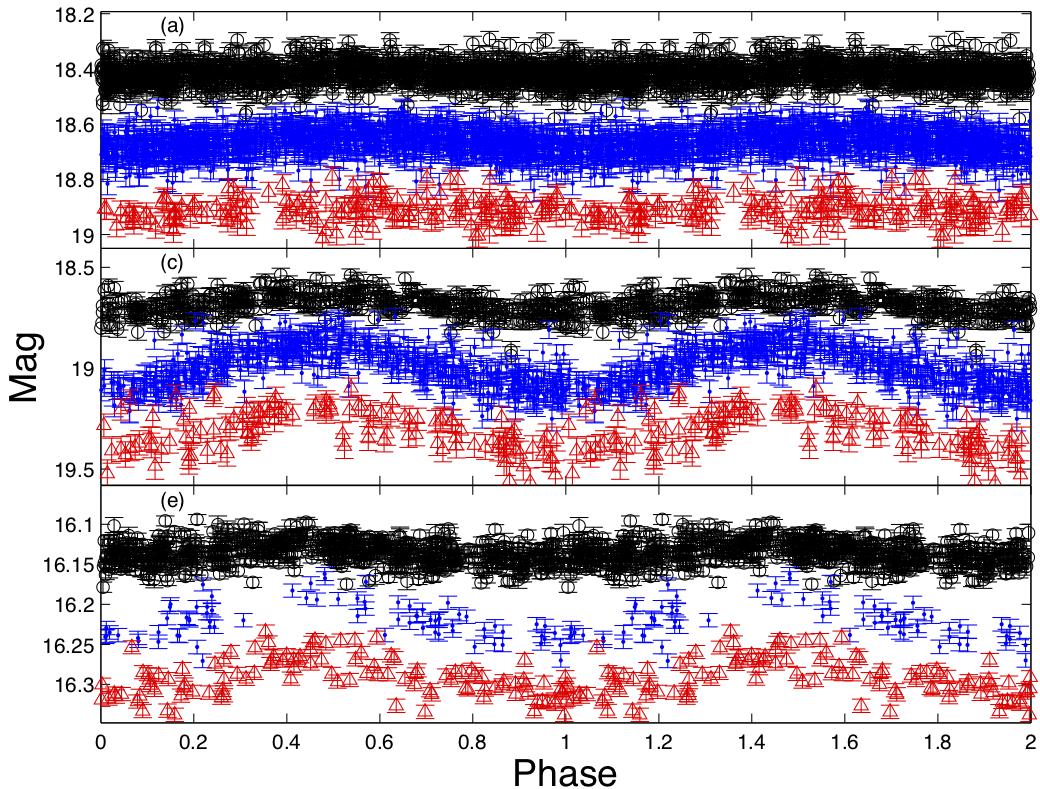}
	\includegraphics[width=8.5cm, height=8.5cm]{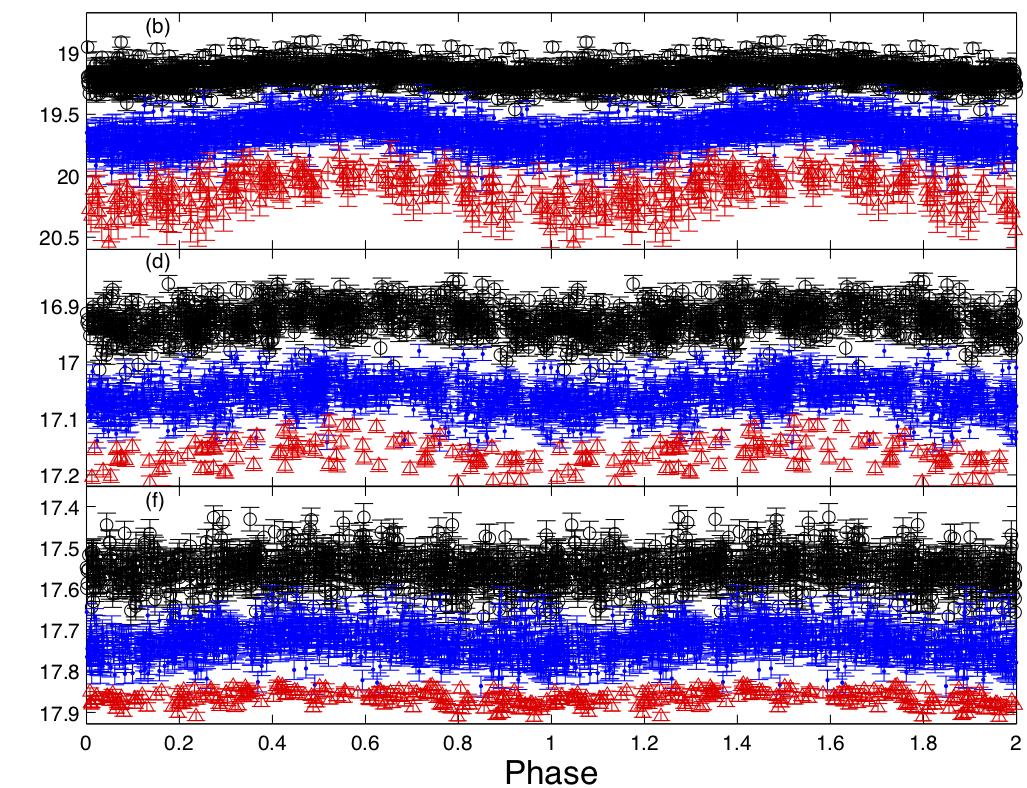}
    \caption{Same as Fig. 5. But for other 6 PCEBs (i.e. SDSS J1638, SDSS J1705, SDSS J2130, SDSS J2208, SDSS J2320, and SDSS J2343).}
    \label{fig:ZTF1638_figure}
\end{figure*}

\subsubsection{SDSS J1638}
SDSS J163824.78+292701.23 (hereafter SDSS J1638) was first classified as a DA+dM binary by \citet{Skiff2009}, then the atmospheric and physical parameters for the components
of this object were derived by \citet{Rebassa2010} and \citet{Liu2012}. It was found that SDSS J1638 is composed of a hot DA WD and a 0.32$M_{\sun}$ M-type dwarf. Thereafter, these parameters were derived by \citet{Ren2018} again based on its spectrum from LAMOST.

The radial velocities for the DA star and M dwarf were determined to be -51.6 and 123.9 \kms respectively by \citet{Morgan2012} who gave an upper limit of 39.49 days for the orbital period of this object based on the RV observations without multi-epoch measurements, suggesting that SDSS J1638 might be a short-period
PCEB and the variability in its luminosity might be detected based on the photometric observations. In order to obtain its actual orbital period for
this object, we collect the photometric observations for it from ZTF DR19, the total of 1028 data points in $g$-band, 1148 data points in $r$-band and 160 data points in $i$-band,
respectively. However, its orbital period can be only derived from the photometric observations in $r$-band and it is 0.454168(16) days (listed in Table~\ref{tab:lc_table}) and the
periodogram indicating its orbital period is also shown in Fig.~\ref{fig:ZTF00321_figure}. As seen from Fig.~\ref{fig:ZTF00321_figure}, there are some possible periods that can not
be ruled out, however. It is found in Table~\ref{tab:chi2_table} that the $\chi^2$-value based on the main peak is also smaller than that based on another peak with a similar power to the main one and a frequency of 3.2045401 ${\rm d}^{-1}$ ($P=0.3120573$ days ), suggesting that the period indicated by the main peak is the orbital period of this object. Then three light curves are plotted in Fig.~\ref{fig:ZTF1638_figure}(a). As seen from Fig.~\ref{fig:ZTF1638_figure}(a), SDSS J1638 shows periodic variation in its luminosity evidently and it should be a short period PCEB. The variability in the luminosity of SDSS J1638 might be caused by the reflection effect or a star spot due to stellar activity of the M-type dwarf. Meanwhile, the orbital obtained for this object in this work is really shorter than its upper limit listed in \citet{Morgan2012}, suggesting that it might be correct.

\subsubsection{SDSS1705} 
SDSS170517.87+334507.61 (hereafter SDSS J1705) was  classified as a DA white dwarf by \citet{Elsenstein2006} based on the spectra from
SDSS DR4. Then it was found to be a binary system containing a WD and a cool companion by \citet{Silvestri2007} and the DA star in this binary system was found to be very hot
one by \citet{Kleinman2013} and \citet{Anguiano2017}. A radial velocity of the DA star was derived to be -47.78 \kms, and the distance
was estimated to be 1685.9 pc away from the the Earth which is consistent with a distance of 2027.6$\pm$864.1 pc indicated by its parallax from Gaia DR3. Another radial velocity of this object was estimated to be 292.08 \kms from the redshift based on its LAMOST spectrum \citep{Zhang2022}, suggesting that this object
might show a large variability in its radial velocity and thus a short period PCEB.

Based on the photometric data from ZTF DR2, the orbital periods for this object were derived to be 0.254066 days in $g$-band and 0.34053 days in $r$-band by \citet{Chen2020},
suggesting that it is necessary to find the true orbital period  for this object because the different periods were given by \citet{Chen2020} based the photometric data from different passbands. The light curves based on the photometric data from ZTF DR19 is analyzed again and only one periodogram implying its orbital period based on $r$-band data is shown in Fig.~\ref{fig:ZTF00321_figure}. Its orbital periods are determined as 0.3405506(97) days in $g$-band, 0.3405451(81) days in $r$-band and 0.3405535(137) days in $i$-band, respectively. Our results are in well agreement with a period determined by \citet{Chen2020} based on ZTF $r$-band data, and a different orbital period derived by \citet{Chen2020} from $g$-band data might be a result of a selection of wrong peak owing to the less data points in $g$-band at that time or the influence of some scattered data points. In fact, it is found in Fig.~\ref{fig:ZTF00321_figure} that a peak with a similar power to the main one and a frequency of 3.9391472 ${\rm d}^{-1}$ (corresponds to an orbital period of 0.2538621 days) is in agreement with the peak which was used to derive its orbital period based on ZTF $g$-band data by \citet{Chen2020}. However, the $\chi^2$-value based on this peak is larger than that based on main one (see Table~\ref{tab:chi2_table}), which implies that the period indicated by the main peak is the orbital period of this object, then its three phase-folded light curves are displayed in Fig.~\ref{fig:ZTF1638_figure}(b).

\subsubsection{SDSS J2130}

SDSS J213019.79+061204.58 (SDSS J2130) was first identified as a WD+MS binary by \citet{Gentile2015} and  \citet{Kepler2015} who gave the atmospheric and physical parameters for the WD in this object as the followings: $T_{\rm eff}=34,131$K, ${\rm log}g=7.730$ and $M_{\rm WD}=0.534 M_{\sun}$. This object was classified as a RR Lyr-type variable with a pulsation period of 0.34116743 days by \citet{Sesar2017} and \citet{Gavras2023}.

Meanwhile, the hydrogen Balmer emission lines are also presented in its SDSS spectrum (see Fig.~\ref{fig:spec_figure} ), and thus suggests that SDSS J2130 might be a short period PCEB. In order to obtain
the orbital period for this object, we collect the photometric data for it from ZTF DR19, then the light curves in $g$, $r$ and $i$-bands are analyzed, however only one periodogram implying its orbital period is shown in Fig.~\ref{fig:ZTF00321_figure} since the periodograms obtained from photometric observations in $g$ and $i$-bands for this object show the similar peak distribution to that derived from $r$-band data. Meanwhile, as the objects mentioned above, the period indicated by the main peak in Fig.~\ref{fig:ZTF00321_figure} should be the orbital period of this object since the $\chi^2$-value based on the main peak is smaller than that based on another peak with a similar power and a frequency of 2.8123848 ${\rm d}^{-1}$, corresponding a period of 0.355574 days (see Table~\ref{tab:chi2_table}), therefore its orbital period is derived
to be 0.2621130(14), 0.2621132(12) and 0.2621115(14) days according to $g$, $r$ and $i$-band photometric data, respectively. Its phase-folded light curves are shown in Fig.~\ref{fig:ZTF1638_figure}(c). As seen from Fig.~\ref{fig:ZTF1638_figure}(c), three light curves show the same trend of change, implying that this object should be a
short period PCEB with an orbital period of about 6.2907 hours. Although we attempt to find a peak with a similar power to the main one and a period matching its known period obtained by \citet{Sesar2017} and \citet{Gavras2023}, however we do not find any peak that can satisfy the requirements from Fig.~\ref{fig:ZTF00321_figure}, suggesting that a different orbital period obtained for this object by \citet{Sesar2017} and \citet{Gavras2023} is not a result of the selection of a wrong peak.}

\subsubsection{SDSS J2208}
SDSS J220849.00+122144.73 (SDSS J2208) was first classified as a WD+MS binary by \citet{Silvestri2007} based on its spectrum from SDSS DR5. The atmospheric and
physical parameters for the DA star in this object were obtained by previous investigators \citep[e.g.][]{Rebassa2010,Morgan2012}. It was found that the DA WD in SDSS J2208 is a massive one and the radial velocities for both components were derived to be 31.10 \kms for the DA star and 11.70 \kms for the M star by \citet{Rebassa2010}. Meanwhile, Balmer
emission lines are presented in its spectrum from SDSS. These observational characteristics indicate that SDSS J2208 might be a short-period binary formed from
common envelope evolution.

An orbital period was estimated to be 0.34 days by \citet{Morgan2012} based on RV measurements without multi-epoch, another different orbital period of 1.903 days was determined 
by \citet{Ritter2003} and \citet{Gavras2023} for this object according to the photometric data from SDSS or ASAS. In order to find the accurate period for
SDSS J2208, we collect the photometric observations for it from ZTF DR19, the total of 655 data points in $g$-band, 925 data points in $r$-band and
96 data points in $i$-band are obtained, then the light curves are analyzed and only one periodogram indicating its orbital period based on $r$-band data is also shown in Fig.~\ref{fig:ZTF00321_figure}. In addition, it is found in Table~\ref{tab:chi2_table} that the $\chi^2$-value based on the main peak is smaller than that based on another peak with a similar power to the main one, implying that the period indicated by the main peak is the orbital period of this object, then its orbital period is derived to be 0.654228(21) days in $g$-band and
0.654242(14) days in $r$-band, respectively. Three light curves based on an orbital period derived from the $r$-band data are displayed in Fig.~\ref{fig:ZTF1638_figure}(d).
It is found in Fig.~\ref{fig:ZTF1638_figure}(d) that this object should be a short-period PCEB. However, the orbital periods obtained for this object by us or \citet{Ritter2003} are
longer than its upper limit listed in \citet{Morgan2012}, this might be caused by its upper limit resulted by the unsuitable RV measurements \citep[$V_{\rm dM}=-323.3$ and $V_{\rm WD}=323.3$ \kms,][]{Morgan2012}. In addition, an orbital period derived for this object by \citet{Ritter2003} and \citet{Gavras2023} might be a result of the selection of a wrong peak. In fact, it is found in Fig.~\ref{fig:ZTF00321_figure} that an period indicated by a peak with a similar power to main one and a frequency of 0.52567394 ${\rm d}^{-1}$ ($P$=1.90232 days) is in agreement with that used to derive period for this object by \citet{Ritter2003} and \citet{Gavras2023}.

\subsubsection{SDSS J2302}
SDSS J230202.50-000930.04 (SDSS J2302) was first identified as DA/M binary based on a spectrum from SDSS DR9 by \citet{Li2014} who gave
the atmospheric and physical parameters for both components of SDSS J2302. Then these parameters were studied again
by \citet{Kepler2015} and it was found to be a DA+M3 binary. Meanwhile, SDSS J2302 was found to be a variable star with an orbital period of 0.9098531 days by
\citet{Ivezic2007}. This implies that SDSS J2302 might be a PCEB.

We collect the photometric data for this binary system from ZTF DR19 to obtain the accurate period, the total of 330 data points in $g$-band, 360 data points in $r$-band
and 71 data points in $i$-band are obtained, respectively, then the light curves of this object are analyzed, then the periodogram indicating its orbital period based on $r$-band data is also shown in Fig.~\ref{fig:ZTF00321_figure}. As the objects mentioned above, the $\chi^2$-values listed in Table~\ref{tab:chi2_table} also suggest that the period indicated by the main peak in Fig.~\ref{fig:ZTF00321_figure} is the orbital period of this object, then its orbital period is derived to be 0.2376133(5) days in
$g$-band and 0.2376198(3) days in $r$-band, respectively.  At last, the three light curves based on an orbital period obtained from $r$-band data are shown
in Fig.~\ref{fig:ZTF0036_figure}(c). It is found in Fig.~\ref{fig:ZTF0036_figure}(c) that SDSS J2302 should be an eclipsing PCEB with an orbital period of
about 5.703 hrs, implying that its orbital periods derived by us are very different from that derived by \citet{Ivezic2007}. We attempt to find out a peak indicating a period that can match its known period, however, we do not find any peak that meet the requirements from Fig.~\ref{fig:ZTF00321_figure}, suggesting that the difference between its known period and
our results is not a result of a wrong peak selection, and might be caused by the limited number of the repeated observations ($\sim$10) adopted by  \citet{Ivezic2007} and thus an eclipsing binary with a much shorter duration than
its orbital period could easily escape detection \citep{Ivezic2007}.

\subsubsection{SDSS J2320}
SDSS232004.02+270623.73 (SDSS J2320) was first identified as a DA white dwarf by \citet{Oswalt1984}, and then was reclassified as a
DA+M binary by \citet{Guo2015}. The atmospheric and physical parameters for the DA star in SDSS J2320 were derived to be $T_{\rm eff}=31,480(491)$ K, ${\rm log}g=7.68(6)$,
$M=0.50M_{\sun}$ and a distance away from the earth about 246 pc by \citet{Limoges2010} based on the spectrum from KISO Schmidt survey. A similar result about these parameters was obtained again according to the spectrum from LAMOST DR5 by \citet{Ren2018} who gave an M4-type dwarf in this object.

The emission lines at Balmer series are presented in its optical spectra from LAMOST \citep{Guo2015}, implying that this object might be a short-period PCEB. Therefore, the variability in its luminosity might be detected, we collect the photometric data from ZTF DR19 to discover whether the luminosity of SDSS J2320 is variable or not. As a result, the total
of 641 data points in $g$-band, 78 data points in $r$-band and 111 data points in $i$-band are found out for it. Then the light curves are analyzed and only one periodogram indicating its orbital period based on $g$-band data is also shown in Fig.~\ref{fig:ZTF00321_figure}. As the objects mentioned above, the $\chi^2$-values listed in Table~\ref{tab:chi2_table} also imply that the period indicated by the main peak should be the orbital period of this object, then its orbital period is derived
to be 0.794569(12) days in $g$-band and 0.794531(10) days in $i$-band, respectively. At last, three light cures based on an orbital period derived from $g$-band data (a more
data point set) are plotted in Fig.~\ref{fig:ZTF1638_figure}(e). As seen from Fig.~\ref{fig:ZTF1638_figure}(e), three light curves exhibit the same trend in its luminosity change, implying that the orbital period obtained in this work is accurate at present.

\begin{figure}
	\includegraphics[width=\columnwidth]{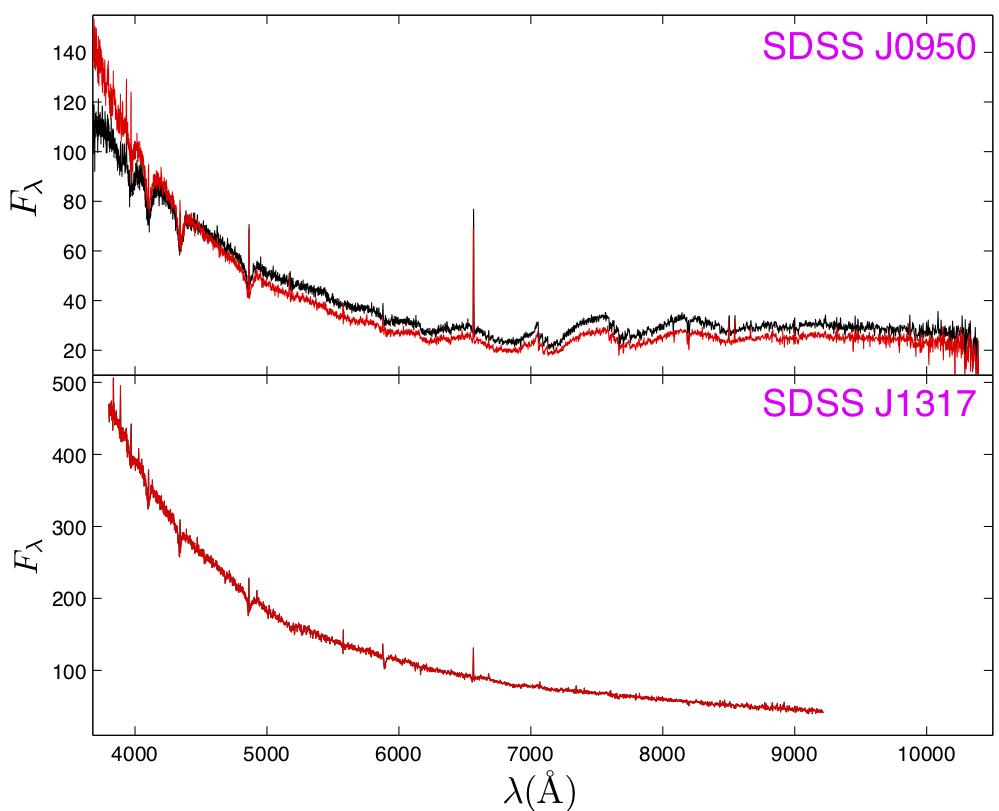}
    \caption{Same as Fig.2, but for two short-period PCEBs (SDSS J09500 and SDSS J1317) with He I emission lines. The different colours indicate the different available spectra and $F_{\rm \lambda}$ in the same units as that in Fig. 2.}
    \label{fig:spec-He_figure}
\end{figure}

\subsubsection{SDSS J2343}
SDSS234312.96+154106.43 (SDSS J2343) was first identified as a WD+MS binary by \citet{Silvestri2007}, and its atmospheric and physical parameters for this object
had been studied by the previous investigators \citep[e.g.][]{Rebassa2010,Morgan2012}. It was found that the DA star in SDSS J2343 is a massive WD with a distance
of 225 pc away from the earth \citep{Rebassa2010}. The radial velocities were derived to be 93.5 \kms and -224.3 \kms for the DA star and M-type star, respectively by
\citet{Morgan2012} who also estimated an upper limit of 6.64 days for orbital period of this object based on several RV measurements, suggesting that SDSS J2343 might be a PCEB.

Since its optical spectrum from SDSS exhibits evident emission features at Balmer series (see Fig.~\ref{fig:spec_figure}). This also indicates that SDSS J2343 might be a short-period PCEB, and thus
the variation in its luminosity might be detected. We collect the photometric data for SDSS J2343 from ZTF DR19, the total of 663 data points in $g$-band, 896 data
points in $r$-band and 171 data points in $i$-band are obtained, respectively. Its light cures are analyzed and only one periodogram indicating its orbital period based on $r$-band data is also shown in Fig.~\ref{fig:ZTF00321_figure}. Although there are some peaks with a similar power to the main peak in Fig.~\ref{fig:ZTF00321_figure}, the period indicated by the main peak should be the orbital period of this object based on the $\chi^2$-values listed in Table~\ref{tab:chi2_table} as the same reason as that for the objects mentioned above, then its orbital period is derived to be 0.5687198(130) days in $g$-band and
0.5687399(106) days in $r$-band respectively and the light curves based on an orbital period determined by the $r$-band observations are shown in Fig.~\ref{fig:ZTF1638_figure}(f).
As seen from Fig.~\ref{fig:ZTF1638_figure}(f), SDSS J2343 is indeed a short-period PCEB and the periodic change in its luminosity might be caused by the reflection
effect or dark spot due to magnetic activity. In addition, the orbital period obtained for this object is indeed shorter than  its upper limit derived by \citet{Morgan2012}, and thus might be correct.

\subsection{Short-period PCEBs with Hydrogen and He I emission features}

\begin{figure}
	\includegraphics[width=\columnwidth]{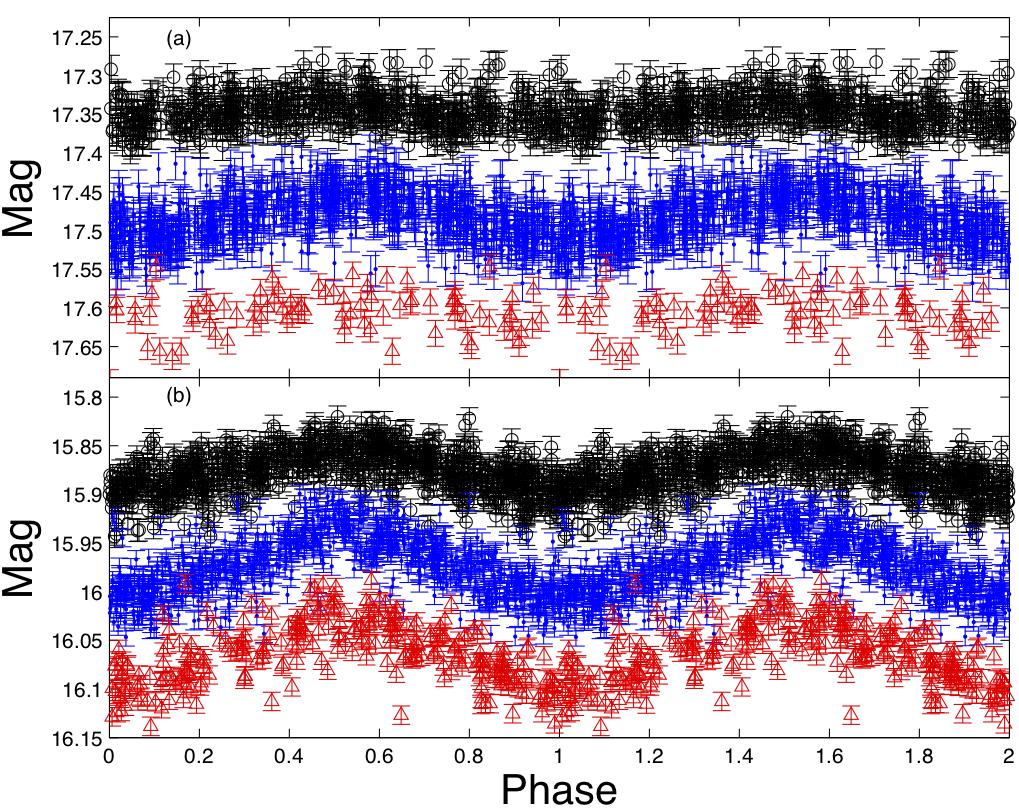}
    \caption{Same as Fig. 5, but for two short-period PCEBs (SDSS J0950 and SDSS J1317) with He I emission feature.}
    \label{fig:ZTF0950_figure}
\end{figure}

\begin{figure*}
	\includegraphics[width=\columnwidth]{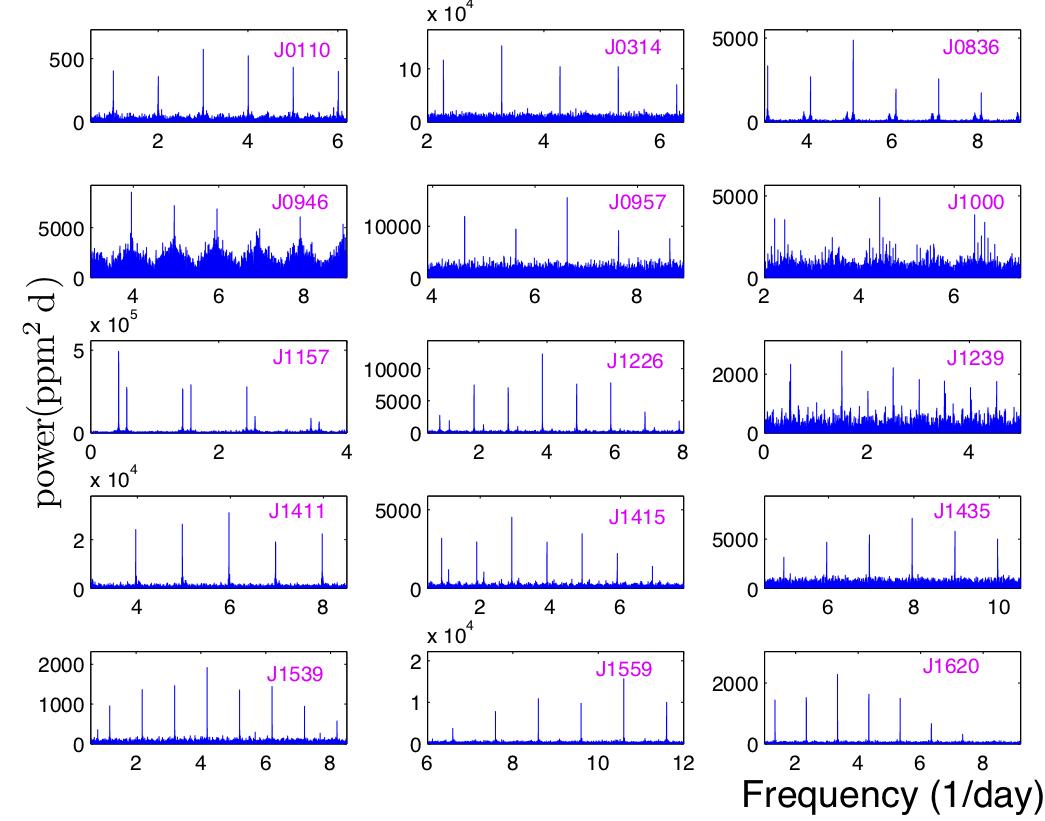}
	\includegraphics[width=\columnwidth]{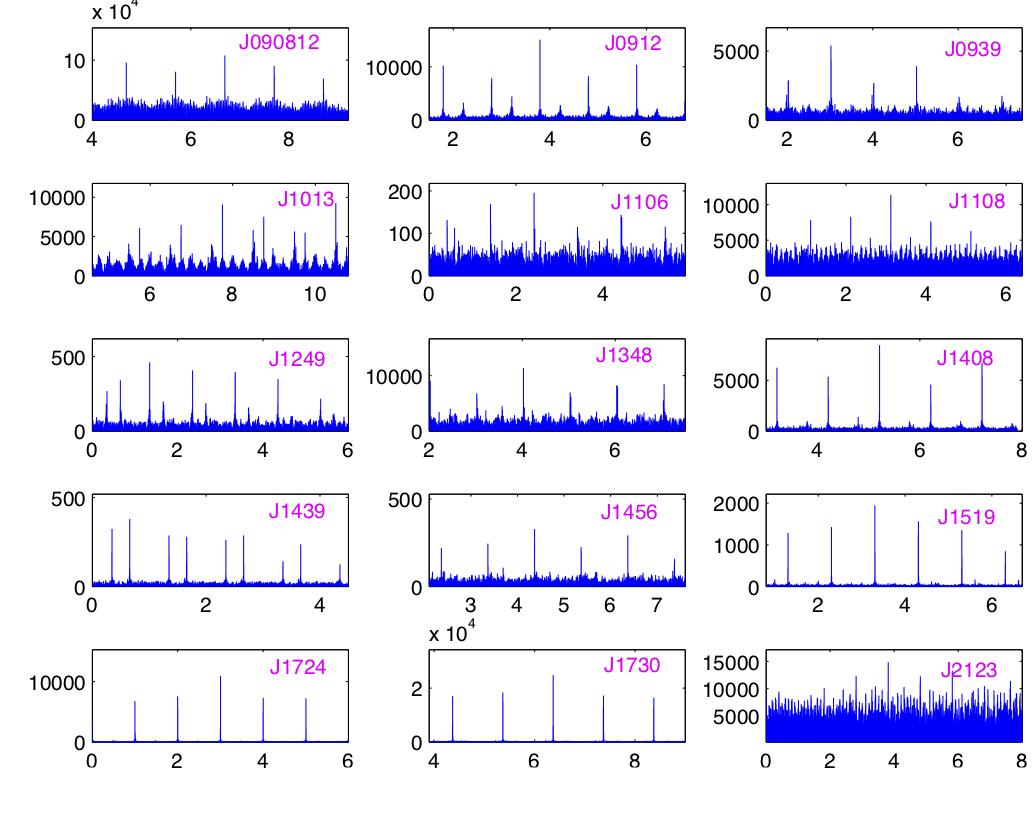}
    \caption{The periodograms indicating orbital periods derived from $r$-band ZTF photometric observations for 30 PCEBs with an accurate orbital period.}
    \label{fig:ZTF0110_figure}
\end{figure*}

\subsubsection{SDSS J0950}
SDSS J095043.94+391541.62 (SDSS J0950) was first identified as a WD+MS binary system by \citet{Kleinman2004}, then its atmospheric and physical parameters were
determined through spectral analysis by many investigators \citep[e.g.][]{Rebassa2010,Tremblay2011, Debes2011, Li2014,Bedard2020}. These results imply
that this binary system should contain a very hot young WD component. Meanwhile, it is found in Fig.~\ref{fig:spec-He_figure} that its SDSS optical spectra display not only the strong emission lines at hydrogen Balmer series, but also the emission lines at He I $\lambda\lambda$5876 and $\lambda\lambda$6681, together with Ca II H,K. These emission lines exhibit a narrow single peak, implying that SDSS J0950 shows the same emission features as LAMOST J143947.62-010606.8 with an orbital period of about 1.522608 days \citep{Gavras2023}.

We collect the photometric data for this object from ZTF DR19, and analyze its light curves in $g$ and $r$-bands after some scattered data points are removed. The result is listed in Table~\ref{tab:lc_table} and only one periodogram implying its orbital period based on $r$-band is shown in Fig.~\ref{fig:ZTF00321_figure}. For the same reasons as those alleged for the objects mentioned above, the period implied by the main peak is the orbital period of this object (see Table~\ref{tab:chi2_table}), then the orbital period of this object is derived to be 1.167186(21) and 1.167341(15) days based on $g$ and $r$-band data, respectively.  The phase of each datapoint in $g$, $r$ and $i$-bands is calculated according to
an orbital period derived from $r$-band data, then its phase-folded light curves are shown in Fig.~\ref{fig:ZTF0950_figure}(a). As seen from Fig.~\ref{fig:ZTF0950_figure}(a),
SDSS J0950 is composed of a hot WD and an M-type dwarf with an orbital period of 1.167341 days, and is therefore a detached binary system, rather than a*
cataclysmic variable.

\subsubsection{SDSS J1317}
SDSS J131751.72+673159.36 (SDSS J1317)  was identified as a cataclysmic variable by \citet{Green1986}. The atmospheric and physical parameters were determined based
on its SDSS spectrum by \citet{Rebassa2010, Tremblay2011} and \citet{Bedard2020}. It was found that this WD/MS binary system should be composed of a hot DA WD and
an M-type dwarf. The radial velocities were determined to be 460.4(22.5) and -449.2(35.8) \kms at different epochs by \citet{Pourbaix2005}, this implies that this object
might be a close binaries.

Meanwhile, it is found in Fig.~\ref{fig:spec-He_figure} that its SDSS optical spectrum displays not only the strong emission lines at hydrogen Balmer series, but also the emission lines at He I $\lambda\lambda$5876 and 6681 and each emission line shows a single peak, the emission features in its spectrum are similar to those in the spectra of BE UMa \citep[with $P_{\rm orb}=2.2909892$ days, ][]{Sanchez2023} and HK Leo \citep[with $P_{\rm orb}=1.7598817$ days,][]{Gavras2023}. In addition, \citet{Holl2023} discovered a spurious signal with a frequency of 0.29568185 ${\rm d^{-1}}$ (corresponding to a period of 3.382013 days) related to time-dependent scan angle for this object according to Gaia DR3 $G$-band time series data.  In order to find  whether this spurious period is the orbital period of this object, we collect the photometric data from ZTF DR19, then analyze the light curves in $g$, $r$ and $i$-bands, the result is listed in Table~\ref{tab:lc_table} and the periodogram implying its orbital period from $r$-band observations is also shown in Fig.~\ref{fig:ZTF00321_figure}. For the same reason as those for SDSS J0029, the period indicated by the main peak in Fig.~\ref{fig:ZTF00321_figure} should be the orbital period of this object (see Table~\ref{tab:chi2_table}), then its orbital
period is derived to be 3.38084(20), 3.38136(32)  and 3.38115(61) days based on $g$, $r$ and $i$-band data, respectively. The phase for each data point in three
passbands is calculated according to a period obtained from $g$-band data, then the phase-folded light curves are displayed in Fig.~\ref{fig:ZTF0950_figure}(b). This suggests that the period discovered for this object by \citet{Holl2023} should be the orbital period rather than the spurious period for SDSS J1317. 

\section{Discussion and conclusions}

The common envelope evolution is one of the most uncertain processes in binary evolution \citep{Politano2007,Nebot2009}. The post common envelope binaries (PCEBs) are the
direct products of common envelope evolution and thus play an important role in understanding common envelope evolution of binaries \citep{Paczynski1976,Taam2000,Nebot2009,Rebassa2007}.
In this work, we attempt to discover some PCEBs from WDMS binaries with emission line(s) identified from LAMOST and/or SDSS based
on the photometric data from ZTF DR19. As a result,
55 PCEBs with an orbital period within a range from 2.2643 to 81.1526 hours are found out based on the photometric data, however most of them had been discovered
by previous investigators. In these short-period PCEBs, 6 of them are newly discovered and the orbital periods of 19 PCEBs have been improved (see the first 25 objects in Table~\ref{tab:lc_table}) based on a match with Simbad database.  A detailed comparison between our results and those obtained by previous investigators is shown in Fig.~\ref{fig:spec_period}. As seen from Fig.~\ref{fig:spec_period}, our results are consistent with their known ones for most of the known short period PCEBs except for 8 PCEBs
(indicated by squares) with the upper limits of their orbital periods \citep{Morgan2012}, implying that the method \citep[named $Period04,$][]{Lenz2005} used in this work is effective although almost all periodograms indicating the orbital periods for these short period PCEBs show several peaks with a similar power, because the similar phenomena also occur in the periodograms for 30 short period PCEBs with an accurate period (see Fig.~\ref{fig:ZTF0110_figure}). A possible explanation for this phenomenon is that  the orbital periods of these binary stars are obtained from the discontinuous observation data
from ZTF, which might produce some spurious signals.

It is found in Fig.~\ref{fig:ZTF00321_figure} and Fig.~\ref{fig:ZTF0110_figure} that there are some peaks with a similar power to the main one and thus some possible periods for these PCEBs can not be directly ruled out. In order to find the orbital periods for the newly discovered or period improved PCEBs, we use a $\chi^2$-method to check the deviation degree of the phase-folded light curves based on two orbital periods indicated by the main peak and another peak with a similar power to the main one from the 'averaged' light curves constructed by 50 normal points. It is found in Table~\ref{tab:chi2_table} that the deviation degree of light curve based on the orbital period indicated by the main peak is lower than that of light curve based on another peak for each of them, suggesting that the periods indicated by the main peaks in their periodograms are their orbital periods, also suggesting that the
method used in this work is effective. In addition, some peaks with a similar power to the main one in their periodograms might be a result of a small amplitude of light variation in these short period PCEBs composed of a WD and a low-mass dwarf, and the limited repeated observations or scattered data points would lead to select a wrong peak for some binaries (such as SDSS J1541, SDSS J1705 and SDSS J2302). Therefore, it is necessary to use the multi-band photometric observations for deriving the orbital periods of short period PCEBs.

The upper limits were estimated for 8 short period PCEBs based on several RV measurements by \citet{Morgan2012}, and thus they are hardly considered as their orbital periods. The orbital periods derived for them (except for SDSS J2208) in this work are shorter than the upper limits listed in \citet{Morgan2012}, and thus the orbital periods obtained in this work based on the photometric data from ZTF DR19 might be correct. In addition, an orbital period of SDSS J2302 \citep[with $P_{\rm orb}=0.9098531$ days,][]{Ivezic2007} is derived to be 0.2376198 days and this object is found to be an eclipsing PCEB in this work. The difference between our result and theirs might be caused by the different observations used. The result obtained for this object by \citet{Ivezic2007} only based on a limited number of repeated observations \citep[$\sim10$,][]{Ivezic2007}. Although a limited number of high-precision observations can reveal the variability of stellar luminosity, it is difficult to obtain their exact periods, so that an eclipsing binary with a much shorter eclipse duration than its orbital period could easily escape detection \citep{Ivezic2007}. 

As seen from Fig.~\ref{fig:spec_figure} and Fig.~\ref{fig:spec-He_figure}, their optical spectra from LAMOST and/or SDSS show the evident emission line(s) at Balmer series or even He I $\lambda\lambda$5876 and $\lambda\lambda$6681.  A possible explanation for this behavior is photoionization and recombination due to irradiation of M dwarfs because of their very hot WD companions with an effective temperature higher than $\sim$10,000 K \citep{Silvestri2006}. Another one is that the emission features might be a result of the magnetic activity of M dwarf components in these PCEBs, since the M dwarfs in PCEBs are younger, and thus more active than the field M dwarfs \citep{Rebassa2013b}.  Therefore, the reflection effect or star spot due to stellar magnetic activity can provide the favorable opportunity for searching the short-period PCEBs from WDMS binaries with emission lines.  Meanwhile, the optical spectra of SDSS J0950 and SDSS J1317 show the characteristic spectra ($\lambda\lambda$8183.27 and $\lambda\lambda$8194.81 absorption doublet) of the M-type dwarfs, suggesting that SDSS J0950 and J1317 should contain an M-type dwarf component. In addition, the emission lines presented in their optical spectra exhibit a single peak, therefore the emission lines in their optical spectra might not be a result of the accretion disk, implying that they are probably not the cataclysmic variables. In fact,
the M-type dwarfs in these two PCEBs with an orbital period longer than 1 days cannot fill their Roche lobes to transfer mass to their WD companion and thus form an accretion disk around these white dwarfs, unless the WDs in them are extremely massive ones.

In addition, we have analyzed the SED of  SDSS J1541 because of the lack of properties of its cool companion. As a result,  it is found that the hot WD SDSS J1541 might be companied by a late-type M dwarf,
it should be produced by common envelope evolution, however, it is still worth further studies how can its thick common envelope be ejected by a very low orbital
energy of the initial binary, because the WD in it is a He-WD with $M \la 0.45 M\sun$, suggesting that most of mass of its MS progenitor
had been ejected by CE evolution, since the oldest globular clusters in the galactic halo are producing $\sim 0.53M_{\sun}$ WDs for MS progenitors with
$M\la 0.8M\sun$ \citep{Kalirai2009,Brown2011}.
Meanwhile, it is found in Fig.~\ref{fig:ZTF0036_figure} that the light curves of three PCEBs (SDSS J0747, SDSS J0908 and SDSS J2302) show the evident eclipses, implying that these PCEBs are close eclipsing binaries with a short period. The eclipsing PCEBs can provide us with the possibility to directly obtain their precise parameters independently of atmospheric parameters. Therefore,  we would observe these eclipsing PCEBs and
analyze their light curves and radial velocity curves to obtain the precise physical parameters for them in the future since their precise parameters are useful to constraint
the CE evolution.

\section*{ACKNOWLEDGEMENTS}
The authors are grateful to the anonymous referee for the valuable suggestions and insightful remarks,
which have improved this work greatly. We also thank prof. D. Koester and P. Bergeron for providing their WD
models. This project was partly supported by the Chinese Natural Science Foundations (Nos. 11773065, 11973081 and 12073070), and
by the Science Research Grants from
National Key R\&D Program of China (2021TYFA1600403) and the International Centre of Supernova, Yunnan Key
Laboratory (No. 202302AN360001).

Funding for the SDSS and SDSS II provided by the Alfred P. Sloan Foundation, the participating
Institutions, the National Science Foundation. The US Department of Energy, the National Aeronautics
and Space Administriction, the Japanese Monbukagakusho and the Max Plank Society, and the Higher Education
Funding Council for England.

This paper makes use of data products of WISE, which is a joint project of the University of California, Los
Angeles, and the Jet Propulsion Laboratory/California Institute of Technology, funded by the National Aeronautics
and Space Administration. this research has made use of the SIMBAD database VizieR service.
\\

\section*{Data Availability}


The data underlying this work will be shared on reasonable requisition to the corresponding author. The various
sky survey data can be available in the public Data Release of ZTF DR19.



\bibliographystyle{mnras}








\bsp	
\label{lastpage}
\end{document}